\documentclass[aps,prd,preprint,showpacs,superscriptaddress,nofootinbib,preprintnumbers]{revtex4}

\usepackage{array}
\usepackage{slashed}
\usepackage{epsf,epsfig}
\usepackage{subfigure}  % use for side-by-side figures
\usepackage{graphicx} % support the \includegraphics command and options
\usepackage{amsmath, amsfonts}
\usepackage{amssymb}

\def\bfnabla{\mbox{\boldmath $\nabla$}}

\def\bfsigma{\mbox{\boldmath $\sigma$}}

\def\bfepsilon{\mbox{\boldmath $\epsilon$}}
\def\em{{\textrm{em}}}
\def\lQ{\Lambda_{\textrm{QCD}}}

\def\als{\alpha_{\textrm s}}

\def\siml{{\ \lower-1.2pt\vbox{\hbox{\rlap{$<$}\lower6pt\vbox{\hbox{$\sim$}}}}\ }} 
\def\simg{{\ \lower-1.2pt\vbox{\hbox{\rlap{$>$}\lower6pt\vbox{\hbox{$\sim$}}}}\ }}

\def\vbfD{{\ \lower-8pt\vbox{\hbox{\rlap{$\!\leftrightarrow$}\lower8pt\vbox{\hbox{$\!\bf D$}}}}\ }} 
\def\dsl{\,\raise.15ex\hbox{/}\mkern-13.5mu D}

\newcommand{\onec}{1\!\!{\rm l}_c}
 
\newcommand{\nn}{\nonumber}
\newcommand{\be}{\begin{equation}} 
\newcommand{\ee}{\end{equation}}
\newcommand{\bea}{\begin{eqnarray}} 
\newcommand{\eea}{\end{eqnarray}}
\newcommand{\beq}{\begin{equation}}
\newcommand{\eeq}{\end{equation}}
\newcommand{\bqa}{\begin{eqnarray}}
\newcommand{\eqa}{\end{eqnarray}}
\newcommand{\bra}[1]{\ensuremath{\langle#1|}}
\newcommand{\ket}[1]{\ensuremath{|#1\rangle}}

\newcommand{\Appendix}[1]%
    {%
     \section{#1}%
      }

\begin{document}
\preprint{\vbox{\halign{ &# \hfil\cr & TUM-EFT 25/11 \cr &\today\cr}}}
\title{Model-independent Study of Electric Dipole Transitions in Quarkonium}

\author{Nora Brambilla}
\affiliation{Physik-Department, Technische Universit\"at M\"unchen,
James-Franck-Str. 1, 85748 Garching, Germany}

\author{Piotr Pietrulewicz}
\affiliation{Institut f\"ur Theoretische Physik, University of Vienna, 
Boltzmanngasse 5, A-1090 Vienna, Austria}

\author{Antonio Vairo}
\affiliation{Physik-Department, Technische Universit\"at M\"unchen,
James-Franck-Str. 1, 85748 Garching, Germany}

\begin{abstract}
The paper contains a systematic, model-independent treatment of electric dipole (E1) transitions 
in heavy quarkonium. Within the effective field theory framework of potential non-relativistic QCD (pNRQCD), 
we derive the complete set of relativistic corrections of relative order $v^2$ 
both for weakly and strongly-coupled quarkonia. 
The result supports and complements former results from potential model calculations.
\end{abstract}


\maketitle
\newpage

\section{Introduction}
In recent years, BES, the B-factories and CLEO have improved on almost all 
heavy-quarkonium radiative transition modes and measured many of them 
for the first time. Part of this impressive progress can be found collected in two reviews 
by the Quarkonium Working Group \cite{Brambilla:2004wf,arXiv:1010.5827}.
Among the most recent developments we mention the first measurements  
of $h_c\rightarrow \eta_c \, \gamma$ by BES \cite{arXiv:1002.0501},
$\Upsilon(2S) \rightarrow \eta_b \, \gamma$ by BABAR \cite{arXiv:0903.1124}, 
$\Upsilon(3S) \rightarrow \eta_b \, \gamma$ by CLEO \cite{arXiv:0909.5474},
$h_b \rightarrow \eta_b \, \gamma$ by BABAR and BELLE \cite{arXiv:1102.4565},
and the  measurements of the branching fractions 
for electromagnetic transitions of the $\chi_b$ states by CLEO \cite{arXiv:1012.0589}
and BABAR \cite{arXiv:1104.5254}. 

Electric dipole (E1) transitions are transitions that change the
orbital angular momentum of the state by one unit, but not its spin. 
Hence, the final state has a different parity and C parity than the initial one. 
Typical E1 transitions are $n^{\prime \,3} P_J \rightarrow n^3 S_1 \, \gamma$, 
e.g.  $\chi_{cJ} \rightarrow J/\psi \, \gamma$ or  
$\chi_{bJ} \rightarrow \Upsilon(1S) \, \gamma$,  
and $n^{\prime \,1}P_1 \rightarrow n^1S_0 \, \gamma$, e.g.  
$h_c \rightarrow \eta _c \, \gamma$ or $h_b \rightarrow \eta _b \, \gamma$. 
E1 transitions happen more frequently than magnetic dipole (M1) transitions.
The branching fraction of E1 transitions can indeed be quite significant for some states, 
e.g. the branching fraction for $\chi_{b1} \rightarrow \Upsilon(1S) \, \gamma$
is   $(35 \pm 8) \%$ and the branching fraction for $\chi_{b2} \rightarrow \Upsilon(1S) \, \gamma$
is   $(22 \pm 4) \%$~\cite{FERMILAB-PUB-10-665-PPD}.
E1 transitions depend on the wave function already at leading order 
and therefore provide a direct insight into the quarkonium state.

Electromagnetic transitions have been treated for a long time by means 
of potential models using non-relativistic reductions 
of phenomenological interactions (see e.g. \cite{PRINT-84-0308}, which will 
be our reference work in the field). A recent comprehensive analysis can be found in \cite{hep-ph/0701208}. 
However the progress made in effective field theories (EFTs) for 
non-relativistic systems, like quarkonium, \cite{hep-ph/0410047}, and 
the new large set of accurate data ask for model-independent analyses. 
M1 transitions have been treated in an EFT framework in \cite{hep-ph/0512369}. 
In this work, we extend that previous analysis to E1 transitions, 
for which a model-independent treatment has been missing so far. 

Effective field theories for quarkonium radiative transitions are built 
on a hierarchy of energy scales \cite{hep-ph/0512369}: the heavy-quark mass $m$, 
the relative momentum of the bound state $p \sim mv$ and the binding energy $E \sim mv^2$, 
where $v\ll 1$ is the heavy-quark velocity in the center of mass frame. 
The condition $v\ll 1$, which works better for bottomonium ($v_b^2 \approx 0.1$) 
than for charmonium ($v_c^2 \approx 0.3$), implies  $ m \gg p \gg E$.
For transitions that involve a change in the principal quantum number, 
the photon energy, $k_\gamma$, scales like $k_{\gamma} \sim mv^2$.
This counting will be assumed throughout the paper, although, 
for transitions between states with the same principal quantum number, 
the photon energy is smaller. A smaller photon energy may be implemented easily  
in the final expressions of the transition widths by suppressing terms proportional 
to $k_\gamma$ accordingly. Observables like the transition widths 
are organized in an expansion in $v^2$; the main purpose of the paper 
is to provide, in an EFT framework, an expression for E1 transition widths 
valid up to relative order $v^2$. The work is partially based on \cite{Piotr}. 

The paper is organized as follows. In Sec.~\ref{secEFT}, 
we construct the EFT Lagrangian relevant for E1 decays up to relative 
order $v^2$. The matching is performed in Sec.~\ref{secMatch}.
In Sec.~\ref{secDecay}, we set up the calculation of the transition rates.
In Sec.~\ref{secRel}, we calculate the relativistic corrections and, 
in Sec.~\ref{secRes}, we write the widths up to order $v^2$.
There we also compare our results with those in \cite{PRINT-84-0308}. 
Finally, in Sec.~\ref{secCon}, we conclude and discuss future applications.

\section{Effective Field Theories}
\label{secEFT}
In this section, we write the low-energy EFT suited to describe 
E1 electromagnetic transitions in heavy quarkonia.
The EFT is constructed in two steps. The first step consists of integrating out 
modes associated with the scale $m$. This leads to 
non-relativistic QCD (NRQCD) coupled to electromagnetism \cite{CLNS-85/641,hep-ph/9407339}.
The second step consists of integrating out modes associated with the scale $mv$, which leads 
to potential non-relativistic QCD (pNRQCD) \cite{hep-ph/9707481,hep-ph/9907240}.
The operators of pNRQCD relevant for M1 transitions have been 
derived in \cite{hep-ph/0512369}; here we consider the electromagnetic operators 
responsible for E1 transitions.

The EFT matrix elements are counted in powers of $v$, while the Wilson coefficients of NRQCD 
are series in $\als$ at the (perturbative) mass scale. 
Heavy quarkonia may be distinguished in weakly-coupled and strongly-coupled quarkonia. 
For weakly-coupled quarkonia, the binding energy is of the order of or larger than 
the typical hadronic scale $\lQ$. This case may be relevant for bottomonium and charmonium 
ground states. Weakly-coupled quarkonia can be treated perturbatively at the   
typical momentum-transfer scale leading to a potential that is Coulombic.
The heavy-quark velocity is of order $\als$ and we may identify the 
momentum-transfer scale with $m\als$ and the binding-energy scale with $m\als^2$.
The Wilson coefficients of weakly-coupled pNRQCD are series in $\als$, which 
can be obtained by expanding in $\als$ when matching pNRQCD to NRQCD. 
We will see, however, that, for E1 operators contributing to the transition widths 
up to relative order $v^2$, such an expansion is not necessary and the obtained 
results will be valid to all orders in the coupling.
In the most general weakly-coupled case, the binding energy is not a perturbative 
scale, $m\als^2 \sim \lQ$, and $\als$ at that scale may not be considered an expansion parameter.

For strongly-coupled quarkonia, $\lQ$ is larger than 
the binding energy  and possibly of the same order as the momentum transfer.
This case may be relevant for all higher bottomonium and charmonium states. 
Since E1 transitions require that at least one involved quarkonium state 
has principal quantum number larger than one, strongly-coupled quarkonia 
are likely to be always involved in such processes. 
Strongly-coupled quarkonia may in general not be treated perturbatively at the 
typical momentum-transfer scale. 
This leads, in particular, to a potential that is not Coulombic.

\subsection{NRQCD}
We consider NRQCD coupled to electromagnetism. The Lagrangian has the form
\begin{equation}
\mathcal{L}_{\textrm{NRQCD}} = \mathcal{L}_{\textrm{2-f}} + \mathcal{L}_{\textrm{4-f}} + \mathcal{L}_{\textrm{light}} \, .
\end{equation}
The term $\mathcal{L}_{\textrm{2-f}}$ denotes the two-fermion sector of the Lagrangian.
If we restrict ourselves to operators relevant for E1 transition widths up to relative order $v^2$,
$\mathcal{L}_{\textrm{2-f}}$ reads
\bea
{\cal L}_{\textrm{2-f}}  &=& 
\psi^\dagger \left( i D_0 + \frac{{\bf D}^2}{2m}  + \frac{{\bf D}^4}{8m^3}\right) \psi
\nn\\
&& 
+ \frac{c_F}{2 m}   \psi^\dagger {\bfsigma} \cdot g {\bf B} \psi
- \frac{c_S}{8 m^2} \psi^\dagger \bfsigma \cdot [-i {\bf D}\times, g{\bf E}] \psi  
+ \frac{c_D}{8m^2}  \psi^\dagger\left[{\bf D} \cdot, g{\bf E} \right]\psi
\nn\\
&& 
+ \frac{c_F^\em}{2 m}   \psi^\dagger \bfsigma \cdot e e_Q {\bf B}^{\em} \psi
- \frac{c_S^\em}{8 m^2} \psi^\dagger \bfsigma \cdot [-i{\bf D}\times, e e_Q {\bf E}^{\em}]\psi  
+ \frac{c_D^\em}{8m^2}  \psi^\dagger\left[{\bf D} \cdot, g{\bf E}^{\em} \right]\psi
\nn \\ 
&& + [\psi \rightarrow i \sigma ^2 \chi^*, A_\mu \rightarrow - A_\mu^T,
      A^{\em}_\mu \rightarrow - A^{\em}_\mu]\,,
\label{NRQCD:Lag}
\eea
where $\psi$ is the Pauli spinor field that annihilates a heavy quark of mass
$m$, flavor $Q$ and electric charge $e e_Q$  ($e_b  = -1/3$, $e_c=2/3$), and 
$\chi$ is the corresponding Pauli spinor that creates a heavy antiquark.
The gauge fields with superscript ``${\em}$'' are the electromagnetic fields, 
the others are gluon fields, $i D_0 = i \partial_0 - g T^a A^a_0 - e e_Q
A^{\em}_0$, $i {\bf D} = i \bfnabla + g T^a {\bf A}^a + e e_Q {\bf A}^{\em}$, 
$[{\bf D} \times, {\bf E}]={\bf D} \times {\bf E} - {\bf E} \times {\bf D}$,  
${\bf E}^i = F^{i0}$, ${\bf B}^i = -\epsilon_{ijk}F^{jk}/2$,  
${\bf E}^{i\,\em} = F^{i0\,\em}$ and  ${\bf B}^{i\,\em} = -\epsilon_{ijk}F^{jk\,\em}/2$. 

The coefficients $c_F$, $c_S$, $c_D$, $c_F^{\em}$, $c_S^{\em}$ and $c_D^{\em}$ are Wilson coefficients of NRQCD. 
Some of them satisfy exact relations dictated by reparameterization (or Poincar\'e) 
invariance \cite{hep-ph/9205228}, e.g. 
\bea
c_S^{\em} &=& 2 c_F^{\em} -1 \,, \qquad  c_S = 2 c_F -1 \,.
\label{rpi:cs}
\eea
The coefficients are one at leading order, but known at least at one loop \cite{hep-ph/9701294}.  
In particular, we have 
\bea
c_F^{\em} \equiv 1+ \kappa_Q^{\em} &=& 1+ C_F \frac{\als}{2\pi} + {\cal O}(\als^2)\,,
\label{kQ} 
\eea
where $C_F = (N_c^2-1)/(2N_c) =4/3$ and $N_c=3$.
The term $\kappa_Q^{\em}$ is usually identified with the anomalous magnetic moment of the heavy quark $Q$;
$\kappa_Q^{\em}$ is less than $0.1$ for charm and bottom.
In general, the Wilson coefficients of NRQCD contain also contributions  
coming from virtual photons of energy or momentum of order $m$. 
These contributions are suppressed by powers of $\alpha$, the fine structure constant, 
and will be neglected in the following.

The terms $\mathcal{L}_{\textrm{4-f}}$ and $\mathcal{L}_{\textrm{light}}$ 
denote respectively the four-fermion sector and the light-field sector of the Lagrangian.
Light fields include light quarks (assumed to be massless), charm quarks 
in the bottomonium case\footnote{
This is the case when the charm mass is of the order of the momentum transfer.
If it is larger, then the charm may be integrated out together with the bottom mass, 
in which case it contributes to the Wilson coefficients of NRQCD 
and does not appear as a light field in the Lagrangian.}, gluons and photons.
$\mathcal{L}_{\textrm{4-f}}$ and $\mathcal{L}_{\textrm{light}}$  contribute to the 
quarkonium potential and wave functions (see \cite{hep-ph/0410047} and references therein), 
but they do not provide new couplings of the heavy quarks with the electromagnetic fields 
relevant for E1 transitions at relative order $v^2$, which is the accuracy we aim at.

\subsection{pNRQCD} 
\label{secpNRQCD}
In NRQCD, degrees of freedom that scale with the momentum transfer and with 
the binding energy are entangled in physical amplitudes, leading to a non-homogeneous power counting. 
These degrees of freedom are disentangled in pNRQCD, 
where degrees of freedom that scale like $mv$ have been integrated out.
The pNRQCD Lagrangian may be decomposed in two terms:
\begin{equation}
\mathcal{L}_{\textrm{pNRQCD}} = \mathcal{L}_{\textrm{0}} + \mathcal{L}_{\gamma\textrm{pNRQCD}} \,.
\end{equation}

The term $\mathcal{L}_{\textrm{0}}$ denotes the part of the pNRQCD Lagrangian that does 
not contain heavy-quark couplings to electromagnetism.
In its gauge-invariant form, it reads
\be
\mathcal{L}_{\textrm{0}} = \int d^3 r \; \mathrm{Tr} \left\{ \mathrm{S}^{\dagger} 
\left( {i {\partial}_0 + \frac{{\boldsymbol \nabla}^2}{4m} +\frac{{\boldsymbol \nabla}_r^2}{m}
+ \frac{{\boldsymbol \nabla}_r^4}{4m^3}} + \dots - V_S \right) \mathrm{S}\right\} 
+ \Delta \mathcal{L}_{\textrm{0}} \,,
\label{Lsinglet}
\ee
where ${\rm S} = S \onec/\sqrt{N_c}$  is a quark-antiquark field that transforms as a singlet 
under $SU(3)_{\rm c}$ and $U(1)_{\em}$. $S$ is labeled by the quark-antiquark distance ${\bf r}$ and depends on the 
center of mass coordinate and time, the derivative $\bfnabla$ acts on the 
center of mass coordinate and the derivative $\bfnabla_r$ on the relative distance  ${\bf r}$.
The dots stand for higher-order kinetic energy terms. Here and in the following, the trace is meant 
over color and spin indices. The Wilson coefficient $V_S$, which is in 
general a function of ${\bf r}$, may be identified with the quark-antiquark color-singlet potential.
It is organized as an expansion in $1/m$, the leading term being the static potential, 
$V_S^{(0)}$. For weakly-coupled quarkonia, $V_S$ may be calculated in perturbation theory, the leading 
term being the Coulomb potential ($V_S^{(0)} = -C_F\als/r$). 
For strongly-coupled quarkonia, $V_S$ follows from a non-perturbative matching to NRQCD.

The term $\Delta \mathcal{L}_{\textrm{0}}$ describes the propagation of all other low-energy degrees of freedom 
besides the quark-antiquark singlet and their strong interactions.
The low-energy degrees of freedom depend on the specific quarkonium under scrutiny.
For weakly-coupled quarkonia, they are, besides the quark-antiquark singlet field,
the quark-antiquark field ${\rm O} = \sqrt{2}\, O^aT^a$, which transforms as an octet under 
$SU(3)_{\rm c}$ and as a singlet under $U(1)_{\em}$, light quarks, low-energy gluons and 
photons; $\Delta \mathcal{L}_{\textrm{0}}$ then reads
\bea
\Delta \mathcal{L}_{\textrm{0}} &=& \int d^3 r \; 
\mathrm{Tr} \Bigg\{ \mathrm{O}^{\dagger} 
\left( {i D_0 + \frac{{\boldsymbol D}^2}{4m} +\frac{{\boldsymbol \nabla}_r^2}{m}} + \dots - V_O \right) \mathrm{O}
\nonumber\\
&& \hspace{2cm}
+  V_A \left(  \mathrm{O}^{\dagger} {\bf r}\cdot g{\bf E}  \mathrm{S} 
+  \mathrm{S}^{\dagger} {\bf r} \cdot g{\bf E}  \mathrm{O}\right) + \dots \Bigg\} 
+ \mathcal{L}_{\textrm{light}} \,,
\label{Loctet}
\eea
where  $iD_0  \mathrm{O}= i\partial _0  \mathrm{O} - g [T^a A_0^a,  \mathrm{O}]$ 
and $i{\bf D}  \mathrm{O} = i\boldsymbol \nabla  \mathrm{O} + g [T^a {\bf A}^a,  \mathrm{O}]$. 
Like $S$, the fields $O^a$ are labeled by the quark-antiquark distance ${\bf r}$ and depend on the 
center of mass coordinate and time. To ensure that gluons and photons are of low-energy 
(i.e. carry energy and momentum lower than the typical momentum transfer in the quark-antiquark system)
gluon and photon fields are multipole expanded in the relative distance ${\bf r}$ and depend only 
on the center of mass coordinate and time.
Hence the Lagrangian is organized as an expansion in ${\bf r}$ and $1/m$ (inherited from NRQCD). 
The dots in \eqref{Loctet} stand for terms that contribute to E1 transitions beyond our accuracy.
The Wilson coefficient $V_O$ may be identified with the  quark-antiquark 
color-octet potential. It is organized as an expansion in $1/m$, the leading term being the static potential, 
$V_O^{(0)}$. At leading order in perturbation theory, $V_O^{(0)} = \als/(2N_cr)$.
The Wilson coefficient $V_A = 1 + {\cal O}(\als^2)$ \cite{Brambilla:2009bi} provides the strength 
of the chromoelectric dipole interaction. 
$\mathcal{L}_{\textrm{light}}$ denotes the light-field sector of the Lagrangian; 
light fields include light quarks (assumed to be massless), gluons and photons.
For strongly-coupled quarkonia, after having integrated out $\lQ$, only degrees of freedom that are 
color singlet are possible \cite{hep-ph/0002250,hep-ph/0009145}. These are, 
besides the quark-antiquark color-singlet field and photons, the Goldstone bosons
associated to the spontaneous breaking of chiral symmetry. The effects of Goldstone bosons 
on E1 transitions go beyond our accuracy and will be neglected. Hence, for 
strongly-coupled quarkonia we set $\Delta \mathcal{L}_{\textrm{0}} = -F_{\mu\nu}^{\em}F^{\mu\nu\,{\em}}/4$, 
whereas all the complication of the non-perturbative treatment goes in the determination of $V_S$.

The term $\mathcal{L}_{\gamma\textrm{pNRQCD}}$ describes the coupling of heavy quark-antiquark pairs 
with low-energy photons, like those responsible for electromagnetic transitions. 
The power counting goes as follows
\be
r \sim 1/mv, \quad  \bfnabla_r \sim mv, \quad \bfnabla \sim mv^2 , k_{\gamma}, \quad
{\bf E}^{\em} , {\bf B}^{\em} \sim k_\gamma^2 \,;
\label{counting}
\ee
in the case of weakly-coupled quarkonia, one has to consider also low-energy gluons 
that scale with $mv^2$ or $k_\gamma$. The leading operator responsible for E1 transitions is the 
electric dipole operator $\mathrm{S}^{\dagger} {\bf r}\cdot ee_Q{\bf E}^{\em}  \mathrm{S}$, while 
operators relevant for E1 transitions at relative order $v^2$ are those suppressed 
by $v^2$ with respect to it. The part of $\mathcal{L}_{\gamma\textrm{pNRQCD}}$ relevant for E1 transitions is 
(operators relevant for M1 transitions have been listed in \cite{hep-ph/0512369})
\begin{widetext}
\begin{align}\label{LagE1}
\mathcal{L}_{\gamma \textrm{pNRQCD}}=  
\int d^3 r \; \mathrm{Tr} & 
\Bigg\{ V^{r\cdot E} \mathrm{S}^{\dagger} {\bf r}\cdot e e_Q {\bf E}^{\em} \mathrm{S}  
\nonumber\\
& + \frac{1}{24} V^{(r\nabla)^2  r \cdot E} \mathrm{S}^{\dagger} 
{\bf r}\cdot [({\bf r} \cdot \bfnabla)^2  e e_Q {\bf E}^{\em}] \mathrm{S} 
\nonumber\\
& + \frac{i}{4m}V^{\nabla \cdot (r \times B)}  \mathrm{S}^{\dagger} 
\{ \bfnabla \cdot , {\bf r} \times e e_Q {\bf B}^{\em} \} \mathrm{S} 
\nonumber\\
& + \frac{i}{12m} V^{\nabla _r \cdot (r \times (r\nabla)B)}  \mathrm{S}^{\dagger} 
\{ \bfnabla_r \cdot , {\bf r} \times  [({\bf r} \cdot \bfnabla) e e_Q {\bf B}^{\em}] \} \mathrm{S} 
\nonumber \\
& + \frac{1}{4m} V^{(r \nabla) \sigma \cdot B} [ \mathrm{S}^{\dagger}, \bfsigma ] 
\cdot [({\bf r} \cdot \bfnabla) e e_Q {\bf B}^{\em}] \mathrm{S} 
\nonumber\\
& - \frac{i}{4m^2}V^{\sigma \cdot(E \times \nabla _r)} [ \mathrm{S}^{\dagger}, \bfsigma] 
\cdot ( e e_Q{\bf E}^{\em} \times \bfnabla_r) \mathrm{S} \Bigg\} 
+ \Delta \mathcal{L}_{\gamma \textrm{pNRQCD}} 
\,.
\end{align}
\end{widetext}
Note that the condition $k_\gamma \ll mv$ guarantees that we can multipole expand 
the electromagnetic fields regardless of the weakly- or strongly-coupled nature of the quarkonia.
On symmetry grounds, more terms than those listed in \eqref{LagE1} are possible.
However, as we will argue in the next section, these are the only ones that get contributions 
from matching with NRQCD. The first line contains the leading electric dipole operator, 
all other operators are suppressed by $v^2$. 
The coefficients $V^{r\cdot E}$, $V^{(r\nabla)^2 r \cdot E}$,  
$V^{\nabla \cdot (r \times B)}$, $V^{\nabla _r \cdot (r \times (r\nabla)B)}$, 
$V^{(r \nabla) \sigma \cdot B}$ and $V^{\sigma \cdot(E \times \nabla _r)}$ are Wilson coefficients 
that will be computed in the next section.

The term $\Delta \mathcal{L}_{\gamma \textrm{pNRQCD}}$ contains the electromagnetic couplings with the 
other low-energy degrees of freedom besides the quark-antiquark singlet. 
For weakly-coupled quarkonia the only relevant term for E1 transitions at relative order $v^2$ is 
\be
\Delta \mathcal{L}_{\gamma \textrm{pNRQCD}} = 
\int d^3 r \; \mathrm{Tr} \left\{ V_O^{r\cdot E} \mathrm{O}^{\dagger} {\bf r}\cdot ee_Q{\bf E}^{\em} \mathrm{O} \right\}\,,
\label{E1octet}
\ee
which is the electric dipole operator for quark-antiquark states in a color-octet configuration.
The corresponding Wilson coefficient is $V_O^{r\cdot E}$. 
For strongly-coupled quarkonia, we may set $\Delta \mathcal{L}_{\gamma \textrm{pNRQCD}} = 0$.

\section{Matching}
\label{secMatch}
We calculate here the Wilson coefficients of $\mathcal{L}_{\gamma\textrm{pNRQCD}}$, 
while the Wilson coefficients of $\mathcal{L}_{\textrm{0}}$ can be found in \cite{hep-ph/0410047} and 
references therein. Calculating the  Wilson coefficients of $\mathcal{L}_{\gamma\textrm{pNRQCD}}$ 
requires to match NRQCD quark-antiquark Green's functions coupled to one external electromagnetic 
field with pNRQCD ones. The electromagnetic field can be multipole expanded. 
The matching can be done order by order in $1/m$ \cite{hep-ph/9701294,hep-ph/0410047}. 
We will perform the matching at leading order in the electromagnetic coupling 
and at all orders in the strong coupling. 
Suitable Green's functions are static Wilson loops with electromagnetic and gluon field insertions.

\subsection{Matching at ${\cal O}(\als^0)$}
\label{tree-level}
Before calculating the Wilson coefficients of $\mathcal{L}_{\gamma\textrm{pNRQCD}}$ to all orders, 
we consider, in the case of weakly-coupled quarkonium, the matching from NRQCD to pNRQCD 
at ${\cal O}(\als^0)$. The calculation can be performed by expanding and redefining 
the fields in the NRQCD Lagrangian. This goes in two steps.
\begin{itemize}
\item[{\it (i)}]{First, one  projects NRQCD on the quark-antiquark Fock space spanned by 
\be
\int d^3x_1\,d^3x_2 \, \psi^{\dagger}(\boldsymbol{x}_1,t)
\varphi(\boldsymbol{x}_1,\boldsymbol{x}_2,t)\chi(\boldsymbol{x}_2,t)\ket{{\textrm{light}}}\,,
\label{twoquarkstate}
\ee
where $\varphi(\boldsymbol{x}_1,\boldsymbol{x}_2,t)$ is a $3 \otimes 3$ tensor
in color space and a $2 \otimes 2$ tensor in spin space, and 
$\ket{{\textrm{light}}}$ is a state that contains an arbitrary number 
of low-energy gluons, photons and light quarks, but no heavy quarks.
}
\item[{\it (ii)}]{Second, one decomposes 
\bea
\varphi(\boldsymbol{x}_1, \boldsymbol{x}_2,t)  
&=& {\rm P}\, \exp \left(ig\int_{x_2}^{x_1} \!\! {\bf A} \cdot d{\bf x} \right) \,{\rm S}^\prime({\bf r},{\bf R},t)
\nn \\
&&
+ {\rm P} \, \exp \left(ig\int_{\bf R}^{x_1} \!\! {\bf A} \cdot d{\bf x} \right) \,{\rm O}^\prime({\bf r},{\bf R},t)\,
\exp\left(ig\int_{x_2}^{\bf R} \!\! {\bf A} \cdot d{\bf x} \right)\!,
\label{Phi:to:S_and_O}
\\
{\rm S}^\prime({\bf r},{\bf R},t) &=& \exp\left( iee_Q\int_{x_2}^{x_1} \!\!{\bf A}^{\em} \cdot d{\bf x} \right)\, {\rm S}({\bf r},{\bf R},t)  \,,
\label{PhiS:to:S_and_O}
\\
{\rm O}^\prime({\bf r},{\bf R},t) &=& \exp\left( iee_Q\int_{x_2}^{x_1} \!\!{\bf A}^{\em} \cdot d{\bf x} \right)\, {\rm O}({\bf r},{\bf R},t)  \,,
\label{PhiO:to:S_and_O}
\eea
where ${\rm P}$ stands for path ordering, ${\bf R} = ({\bf x}_1+{\bf x}_2)/2$ is the center of mass 
coordinate, and ${\bf r} = {\bf x}_1-{\bf x}_2$. 
The decomposition ensures the gauge invariance of the pNRQCD operators.
Finally, all electromagnetic fields are multipole expanded in ${\boldsymbol{r}}$.
}
\end{itemize}
The resulting ${\cal O}(\als^0)$ expression of the Wilson coefficients is\footnote{
${\cal O}(\als^0)$ refers to the matching between NRQCD and pNRQCD: 
the Wilson coefficients of NRQCD are kept unexpanded.} 
\begin{eqnarray}
V^{r\cdot E} = V_O^{r\cdot E} = V^{(r\nabla)^2 r \cdot E} = V^{\nabla \cdot (r \times B)}  
= V^{\nabla_r \cdot (r \times (r\nabla)B)}  & = & 1 \, ,
\\
V^{(r \nabla) \sigma \cdot B} & = & c_F^{\em} \, , 
\\
V^{\sigma \cdot(E \times \nabla_r)} & = & c_S^{\em} \,. 
\end{eqnarray}
Surprisingly, these relations will turn out to be valid to all orders.

\subsection{Matching photons coupled to light quarks}
\label{beyond}
Photons may couple to heavy quarks or to light quarks.
If we treat the $u$, $d$ and $s$ quarks as massless, then 
the QCD Lagrangian is $SU(3)$-flavor symmetric and the three quarks 
only differ in the electric charges. Since the sum of the three charges 
vanishes, so does the sum of all diagrams where the photon couples to 
the three massless quarks. 

\begin{figure}
  \centering
  \epsfig{file=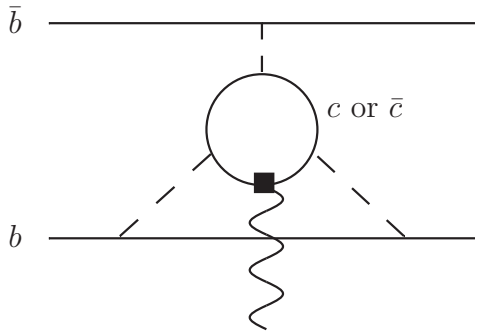,width=0.4 \linewidth,clip=}
  \caption{Leading-order diagram with a $c$-quark loop coupled to an external photon in NRQCD. 
           Dashed lines stand for longitudinal gluons.}
  \label{cquark}
\end{figure}

In the bottomonium case, if the charm-quark mass is of the same 
order as the typical momentum transfer in the system then it should 
be integrated out with that scale. Indeed it contributes 
to the potential \cite{hep-ph/0005066}, although in the spectrum 
it appears to decouple \cite{hep-ph/0108084}. The leading-order 
diagram with a $c$-quark loop coupled to an external photon in NRQCD 
is shown in Fig.~\ref{cquark}. It is of order $\als^3(m_c) \sim v^3$, 
thus beyond our accuracy.

\subsection{Matching of reducible diagrams}
\label{secreducible}
It is useful to observe that certain classes of diagrams in NRQCD contribute just to reducible diagrams in pNRQCD, 
i.e. tree-level diagrams made of potentials and electromagnetic operators of lower order.
Hence, they do not contribute to the matching of new operators. 
This happens whenever a gluonic or electromagnetic contribution can be facto\-ri\-zed out.
In the following, we will illustrate some of these cases.

\begin{itemize}
\item[{\it (i)}]{Consider static NRQCD. It describes the propagation of a static quark located in ${\bf x}_1$
and a static antiquark located in ${\bf x}_2$.
We call $\ket{n}$, states that contain an arbitrary number 
of low-energy gluons and light quarks, but no heavy quarks.
Furthermore we assume that  $\psi^\dagger({\bf x}_1) \chi({\bf x}_2)\ket{n}$ 
is an eigenstate of the static NRQCD Hamiltonian with eigenvalue $E_n$. 
The eigenstates are normalized such that  $\langle n|n\rangle =1$.
In general, both $\ket{n}$ and $E_n$ depend on ${\bf r}$.
}
\item[{\it (ii)}]{The matching between NRQCD and pNRQCD follows 
from equating in the large time limit the time-ordered NRQCD amplitude 
\bea
&& \hspace{-4mm}
\bra{0} \psi({\bf x}_1)\chi^\dagger ({\bf x}_2)
\int d^4 z_1 \left( \psi^\dagger(z_1) O_1(z_1)\psi(z_1) +  \textrm{c.c.}\right) \cdots 
\nonumber\\
&& \hspace{1.5cm}
\times 
\int d^4 z_n \left( \psi^\dagger(z_n) O_n(z_n)\psi(z_n)  +  \textrm{c.c.}\right)\;
\psi^\dagger({\bf x}^\prime_1) \chi({\bf x}^\prime_2)\ket{0} 
\nonumber\\
&&\hspace{-4mm}
= \bra{0} 
\int dt_1 \left(O_1(t_1) + \textrm{c.c.}\right) \cdots 
\int dt_n \left(O_n(t_n) + \textrm{c.c.}\right) 
\nonumber\\
&& \hspace{1.5cm}
\times 
\theta(t_1-t_2)\cdots \theta(t_{n-1}-t_n)
\ket{0}
\,\delta^3({\bf x}_1-{\bf x}^\prime_1)\,\delta^3({\bf x}_2-{\bf x}^\prime_2),
\nonumber
\eea 
with the corresponding pNRQCD one.  
The NRQCD amplitude may be identified in the large time limit with 
a static Wilson loop with $O_1$, ..., $O_n$ and c.c. field insertions \cite{hep-ph/0002250,hep-ph/0009145}.
The operators $O_n(t)$ are gluonic and electromagnetic operators in the Heisenberg representation, 
$\textrm{c.c.}$ stands for charge conjugation (after the equality, it stands for charge conjugation 
and ${\bf x}_1 \leftrightarrow {\bf x}_2$ exchange).
The corresponding pNRQCD amplitude follows from 
$\psi^\dagger({\bf x}_1) \chi({\bf x}_2)\ket{0} \rightarrow S^\dagger({\bf r},{\bf R})|\textrm{vac}\rangle$ 
and $E_0 = V_S^{(0)}$, where  $ |\textrm{vac}\rangle$ is the vacuum of pNRQCD
normalized such that  $\langle \textrm{vac}|\textrm{vac}\rangle =1$, see \cite{hep-ph/0002250}.
}
\item[{\it (iii)}]{
In particular, suppose that the non-electromagnetic operator $\psi^\dagger O \psi + \textrm{c.c.}$ of NRQCD
matches the operator $\displaystyle \int d^3r''\, S^\dagger \delta V S$ of pNRQCD,
then the matching condition reads
\bea
&& \hspace{-4mm}
\bra{0} 
\int dt\, \left(O({\bf x}_1,t) + O^c({\bf x}_2,t)\right) \ket{0}\,
\delta^3({\bf x}_1-{\bf x}^\prime_1)\,\delta^3({\bf x}_2-{\bf x}^\prime_2) 
\nonumber\\
&& \hspace{4mm}
= \langle \textrm{vac}| S({\bf r},{\bf R}) \, \int dt \int d^3R'' \int d^3r'' \,
S^\dagger \delta V S \;\; S^\dagger({\bf r}^\prime,{\bf R}^\prime)|\textrm{vac}\rangle,
\nonumber
\eea
where $O^c$ is the charge conjugated of $O$.
This implies
\bea
\int dt\,e^{iE_0t} \, \bra{0}  O({\bf x}_1) + O^c({\bf x}_2) \ket{0}\,  e^{-iE_0t} =
\int dt\,e^{iV_S^{(0)}t} \,\delta V\,  e^{-iV_S^{(0)}t}, 
\nonumber
\eea
and eventually
\be
\bra{0}  O({\bf x}_1) + O^c({\bf x}_2) \ket{0} = \delta V\,.
\label{coha}
\ee

Another example is the case of a NRQCD amplitude with insertions of two non-electromagnetic 
operators $\psi^\dagger O_1 \psi$  and $\psi^\dagger O_2 \psi$ with their charge-conjugated partners. 
Such an amplitude matches the pNRQCD reducible amplitude made of  
two insertions of $\delta V_1 = \bra{0}  O_1 + O^c_1 \ket{0}$ and 
$\delta V_2 = \bra{0}  O_2 + O^c_2 \ket{0}$, and the pNRQCD amplitude associated with a new potential 
$\delta V_{12}$,
\be
\delta V_{12} = 
-i \int_0^{\infty} dt\, \sum_{n\neq 0} \bra{0} O_1 + O^c_1  \ket{n} e^{-iE_nt}\bra{n}O_2 + O_2^c \ket{0}\, e^{iE_0t} 
+ (1\leftrightarrow 2)\,;
\ee
$n\neq 0$ intermediate state contributions that are exponentially suppressed at large times have been neglected.
}
\item[{\it (iv)}]{Consider now an electromagnetic operator of NRQCD, 
$\psi^\dagger O^{\em}  \psi + \textrm{c.c.}$, that 
matches the operator $\displaystyle \int d^3r''\, S^\dagger {\cal P}^{\em} S$ of pNRQCD.
If  $O^{\em}$ commutes with the gluon fields (at least to the order in the power 
counting we are interested in), then $O^{\em}$ commutes with the static NRQCD Hamiltonian 
and its eigenstates (e.g. $[ O^{\em},\ket{0}] =0$).
The argument of paragraph {\it (iii)} then implies
\be
O^{\em} +  O^{\em\,c} =  {\cal P}^{\em} \,.
\label{coem}
\ee
}
\item[{\it (v)}]{A simple extension of the previous case is the one of 
a NRQCD amplitude with $O_1$, ..., $O_n$, $O^{\em}$ field insertions under the 
condition that $O^{\em}$ commutes with all gluonic operators  $O_1$, ..., $O_n$
(at the order in the power counting we are interested in). 
The amplitude is then proportional to 
\bea
&& O_1(t_1)O_2(t_2)\dots O_n(t_n)O^{\em}(t_{\em})
\Big(\theta(t_{\em}-t_1)\theta(t_1-t_2)\dots\theta(t_{n-1}-t_n)
\nonumber\\
&& \hspace{5.5cm}
+\theta(t_1-t_{\em})\theta(t_{\em}-t_2)\dots\theta(t_{n-1}-t_n)
+ \dots 
\nonumber\\
&& \hspace{5.5cm}
+\theta(t_1-t_2)\dots \theta(t_{n-1}-t_n)\theta(t_n-t_{\em})
\Big)
\nonumber\\
&& = O_1(t_1)O_2(t_2)\dots O_n(t_n) \theta(t_1-t_2)\dots \theta(t_{n-1}-t_n) \; O^{\em}(t_{\em}), 
\eea
where the equality follows from the recursive use of 
$\theta(t_{\em}-t')\theta(t'-t'') + \theta(t'-t_{\em})\theta(t_{\em}-t'') = \theta(t'-t'')\theta(t_{\em}-t'')$.
The equality implies that the amplitude reduces to the product of a pure gluonic amplitude and 
the electromagnetic vertex $O^{\em}$. Therefore it matches the reducible diagram of pNRQCD made of some 
potentials (those that match the non-electromagnetic NRQCD amplitude with $O_1$, ..., $O_n$ field insertions)
and the electromagnetic vertex ${\cal P}^{\em}$.

\begin{figure}
\centering
\epsfig{file=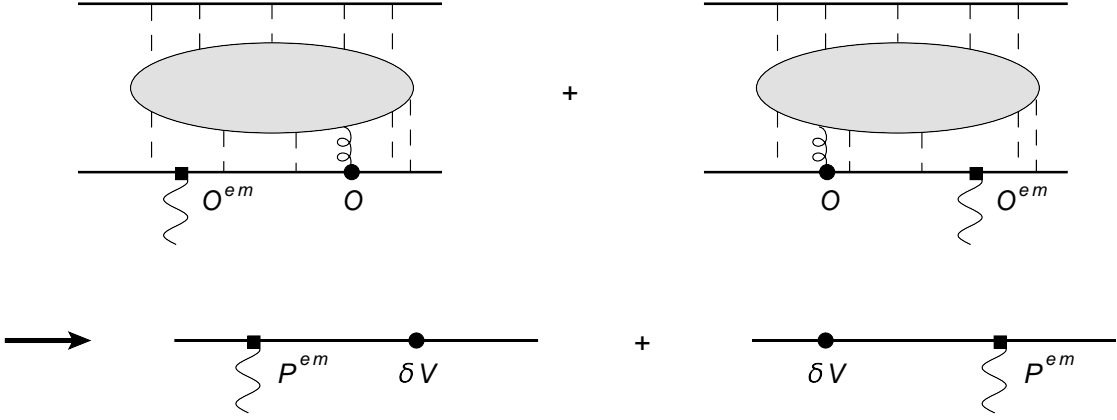,width=0.9 \linewidth,clip=}
\caption{Matching of the amplitude {\it (vi)}: before the arrow are the NRQCD diagrams, after the arrow 
the pNRQCD ones. The continuous line in the pNRQCD diagrams stands for the quark-antiquark singlet 
propagator.}
\label{figred}
\end{figure}

}
\item[{\it (vi)}]{We match now the following amplitude of NRQCD:
\bea
&& \hspace{-9mm}
\bra{0} 
\int dt_1\, \int dt_2\, 
\left(O({\bf x}_1,t_1) + \textrm{c.c}\right)\left(O^{\em}({\bf x}_1,t_2) + \textrm{c.c}\right)
\theta(t_1-t_2)
\nonumber\\
&& \hspace{-9mm}
+\left(O^{\em}({\bf x}_1,t_1) + \textrm{c.c}\right)\left(O({\bf x}_1,t_2) + \textrm{c.c}\right)
\theta(t_1-t_2)
\ket{0}\,
\delta^3({\bf x}_1-{\bf x}^\prime_1)\,\delta^3({\bf x}_2-{\bf x}^\prime_2) , 
\label{redNRQCD}
\eea
under the general assumption that $O^{\em}$ does not commute with $O$.
By inserting a complete set of eigenstates and making use of the fact that 
$O^{\em}$ commutes with the gluon fields, see paragraph {\it{(iv)}}, the amplitude becomes 
\bea
&& \hspace{-4mm}
\int dt_1\, \int dt_2\, \theta(t_1-t_2)\,
e^{iE_0t_1}\Bigg\{
\bra{0} O({\bf x}_1) + O^c({\bf x}_2) \ket{0}
e^{-iE_0(t_1-t_2)} \left(O^{\em}({\bf x}_1) + O^{\em\,c}({\bf x}_2) \right)
\nonumber\\
&& \hspace{1cm}
+\left(O^{\em}({\bf x}_1) + O^{\em\,c}({\bf x}_2)\right)
e^{-iE_0(t_1-t_2)} 
\bra{0} O({\bf x}_1) + O^c({\bf x}_2)\ket{0}
\Bigg\}
\nonumber\\
&& \hspace{6.7cm}
\times
e^{-iE_0t_2}\,\delta^3({\bf x}_1-{\bf x}^\prime_1)\,\delta^3({\bf x}_2-{\bf x}^\prime_2) \,,
\nonumber
\eea
which, according to \eqref{coha} and \eqref{coem}, matches the reducible pNRQCD amplitude
\bea
&& \hspace{-4mm}
\int dt_1\, \int dt_2\, \theta(t_1-t_2)\, 
e^{iV_S^{(0)}t_1}\Bigg\{ \delta V\, e^{-iV_S^{(0)}(t_1-t_2)} {\cal P}^{\em} 
+{\cal P}^{\em} \, e^{-iV_S^{(0)}(t_1-t_2)}  \delta V\, \Bigg\}
\nonumber\\
&& \hspace{6cm}
\times
e^{-iV_S^{(0)}t_2}\,
\delta^3({\bf x}_1-{\bf x}^\prime_1)\,\delta^3({\bf x}_2-{\bf x}^\prime_2) \,.
\nonumber
\eea
Graphically this is shown in Fig.~\ref{figred}.
Therefore an amplitude like Eq.~\eqref{redNRQCD} cancels in the matching against 
a reducible amplitude of pNRQCD. 
}
\item[{\it (vii)}]{Finally, let us consider the case of a NRQCD amplitude with 
two gluonic operators $O_1$ and $O_2$ and corresponding charge-conjugated operators, 
and an electromagnetic operator $O^{\em}$ that commutes with one of the gluonic operators, say $O_2$. 
The amplitude reads
\bea
&& \hspace{-4.2mm}
\int dt_1\, \int dt_2\, \int dt_3\, \theta(t_1-t_2) \, \theta(t_1-t_3)\, e^{iE_0 t_1}
\nonumber 
\\
&& \hspace{-4.2mm}
\times \!
\sum_n
\Bigg\{\!
\bra{0} O_1({\bf x}_1) + O^c_1({\bf x}_2) \ket{n} e^{-iE_n(t_1-t_2)} 
\bra{n} O_2({\bf x}_1) + O^c_2({\bf x}_2) \ket{0} 
\left(O^{\em}({\bf x}_1) + O^{\em\,c}({\bf x}_2) \right)
\!\!\Bigg\}
\nonumber 
\\
&& \hspace{67mm}
\times e^{-iE_0 t_2}\,\delta^3({\bf x}_1-{\bf x}^\prime_1)\,\delta^3({\bf x}_2-{\bf x}^\prime_2) 
\nonumber\\
&& \hspace{-4.2mm}
+ \int dt_1\, \int dt_2\, \int dt_3\, \theta(t_1-t_2)\,\theta(t_3-t_1)\,e^{ iE_0 t_1}
\nonumber 
\\
&& \hspace{-4.2mm}
\times \!
\sum_n
\Bigg\{\!
\left(O^{\em}({\bf x}_1) + O^{\em\,c}({\bf x}_2) \right)
\bra{0} O_1({\bf x}_1) + O^c_1({\bf x}_2) \ket{n} e^{-iE_n(t_1-t_2)} 
\bra{n} O_2({\bf x}_1) + O^c_2({\bf x}_2) \ket{0} 
\!\!\Bigg\}
\nonumber 
\\
&& \hspace{67mm}
\times e^{-iE_0 t_2}\,\delta^3({\bf x}_1-{\bf x}^\prime_1)\,\delta^3({\bf x}_2-{\bf x}^\prime_2) 
\nonumber\\
&& \hspace{-4.2mm}
+\, (1\leftrightarrow 2)\,.
\eea
The $n=0$ component of the amplitude matches the reducible pNRQCD diagrams 
made of insertions of the potentials $\delta V_1$, $\delta V_2$ 
and of the electromagnetic vertex ${\cal P}^{\em}$. 
The sum of the $n\neq0$ components matches 
\bea
&& \hspace{-4mm}
\int dt_1\, \int dt_2\,   \theta(t_1-t_2)\,
e^{iV_S^{(0)} t_1}\Bigg\{ i \delta V_{12}\, e^{-iV_S^{(0)}(t_1-t_2)} {\cal P}^{\em} 
+{\cal P}^{\em} \, e^{-iV_S^{(0)}(t_1-t_2)}  i \delta V_{12}\, \Bigg\}
\nonumber\\
&& \hspace{6cm}
\times
e^{-iV_S^{(0)}t_2}\,
\delta^3({\bf x}_1-{\bf x}^\prime_1)\,\delta^3({\bf x}_2-{\bf x}^\prime_2) \,.
\nonumber
\eea
Therefore, also this kind of amplitude does not induce new operators in pNRQCD.
}
\end{itemize}

\subsection{Matching at $\mathcal{O}(1/m^0)$}
Amplitudes that contribute to the $\mathcal{O}(1/m^0)$ matching of $\mathcal{L}_{\gamma\textrm{pNRQCD}}$
may contain the NRQCD operators (see Eq. \eqref{NRQCD:Lag})
\begin{equation}
-\psi^{\dagger} e e_Q A_0^{\em}\psi, \quad -\psi^\dagger g A_0 \psi \,,
\end{equation}
and the corresponding c.c. ones. Although they just contain one electromagnetic operator, they may 
contain an arbitrary number of longitudinal gluons for they are not suppressed by any power of $1/m$.
The electromagnetic field $A_0^{\em}$ commutes with the gluon fields. 
Therefore it satisfies the condition of Sec.~\ref{secreducible}, paragraph {\it{(iv)}}, and  
the matching condition is given by Eq. \eqref{coem}.
It tells that electromagnetic operators of order $1/m^0$ do not get QCD corrections. 
Hence all $\mathcal{O}(1/m^0)$ Wilson coefficients of $\mathcal{L}_{\gamma\textrm{pNRQCD}}$ 
are fixed at their ${\cal O}(\als^0)$ value:
\begin{equation}
V^{r\cdot E} = V^{(r\nabla)^2 r \cdot E} = 1 \, .
\label{matchsing0}
\end{equation}

For weakly-coupled quarkonia, also the quark-antiquark color-octet sector is relevant.
In this case, the matching is performed by considering NRQCD amplitudes between initial and final states 
that match the color octet state in pNRQCD \cite{hep-ph/9907240}.
Since the electromagnetic field  commutes also with these states, the result of the matching 
to all orders is 
\begin{equation}
V_O^{r\cdot E} = 1 \, .
\label{matchoct0}
\end{equation}

Finally, we observe that, while our argument fixes $V^{r\cdot E}$ to all orders, the same argument does not apply 
to the Wilson coefficient $V_A$ of the chromoelectric dipole operator in Eq. \eqref{Loctet}.
The reason is that the field $A_0$ does not commute with the gluon fields.

\subsection{Matching at $\mathcal{O}(1/m)$} 
\label{matching-1/m}
Amplitudes that contribute to the $\mathcal{O}(1/m)$ matching of $\mathcal{L}_{\gamma\textrm{pNRQCD}}$
contain, besides an arbitrary number of operators of order $1/m^0$, one of 
the NRQCD operators of order $1/m$ (see Eq. \eqref{NRQCD:Lag}):
\begin{equation}
\psi ^{\dagger} \frac{{\bf D}^2}{2m} \psi,\quad
\frac{c_F}{2m}\psi ^{\dagger} \boldsymbol \sigma \cdot g {\bf B} \, \psi, \quad
\frac{c_F^{\em}}{2m}\psi ^{\dagger} \boldsymbol \sigma \cdot ee_Q{\bf B}^{\em} \psi, 
\end{equation}
and the corresponding c.c. ones.
We call the first operator the kinetic energy operator, the second one the 
chromomagnetic dipole operator and the third one the magnetic dipole operator. 
The $\mathcal{O}(1/m)$ amplitudes fall in one of the following categories.

1. The photon is coupled to the magnetic dipole operator. 
This kind of diagrams matches spin-dependent operators and 
may contribute to $V^{(r \nabla) \sigma \cdot B}$.
Using the same argument as for the $1/m^0$ matching,
the amplitude factorizes. Therefore, it holds to all orders that 
\begin{equation}
V^{(r \nabla) \sigma \cdot B} = c_F^{\em} \, ,
\end{equation}
and no other operator is generated.

\begin{figure}[ht]
  \centering
  \epsfig{file=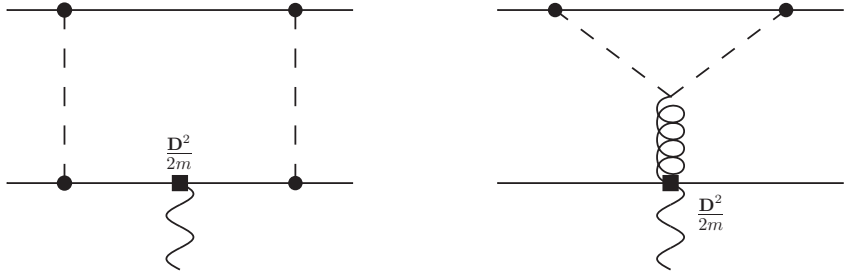,width=0.7\linewidth,clip=}
  \caption{Examples of diagrams at order $1/m$, where the electromagnetic coupling is contained in a covariant derivative.}
  \label{Derivative_diagram1}
\end{figure}

2. The photon is coupled to the kinetic energy operator 
(see e.g. Fig.~\ref{Derivative_diagram1}).
This kind of diagrams matches only operators with magnetic fields,  
hence they may contribute to $V^{\nabla \cdot (r \times B)}$  and $V^{\nabla _r \cdot (r \times (r\nabla)B)}$.
Since the electromagnetic coupling is embedded in a covariant derivative operator,
\begin{equation}
\psi ^{\dagger} (\boldsymbol \nabla - ie e_Q {\bf A}^{\em}) \psi + \textrm{c.c.} \, ,
\end{equation}
this implies that diagrams involving one kinetic energy operator match pNRQCD operators
of the form\footnote{
For the purpose of the present discussion, we do not consider operators 
that do not depend neither on the momentum nor on the electromagnetic field.
They contribute to the $1/m$ potential and have been analyzed in \cite{hep-ph/0002250}. 
}
\bea
&& \int d^3r \; S^{\prime\,\dagger} \,\left( \left\{ \frac{{\delta V_a}}{m},
\left(\bfnabla_{x_1}  - ie e_Q {\bf A}^{\em}({\bf x}_1,t) \right)^2
\right\} + \textrm{c.c.}\right)  S^{\prime} 
\nonumber\\
&&
+ \int d^3r \; S^{\prime\,\dagger} \,\left( \left\{ \frac{{\delta V_b}}{m}{\bf r}\cdot,
i\left(\bfnabla_{x_1}  - ie e_Q {\bf A}^{\em}({\bf x}_1,t) \right)
\right\} + \textrm{c.c.}\right)  S^{\prime} 
\,.
\eea
After the field redefinition \eqref{PhiS:to:S_and_O} and having multipole expanded the electromagnetic 
fields up to include order $r^2$ contributions, the pNRQCD operators may be rewritten as 
\bea
&& \hspace{-5mm}
\int d^3r \; S^{\dagger} \,\left\{ \frac{{\delta V_a}}{m},
\frac{\bfnabla^2}{2} + 2  \bfnabla_r^2 
+2 \left\{ i\bfnabla, \frac{{\bf r}\times ee_Q{\bf B}^{\em}}{4} \right\}
+2 \left\{ i\bfnabla_r,  \frac{{\bf r}\times  [({\bf r} \cdot \bfnabla) e e_Q {\bf B}^{\em}]}{12} \right\}
\right\}S
\nonumber\\
&& \hspace{-5mm}
+  \int d^3r \; S^{\dagger} \, \left\{ \frac{{\delta V_b}}{m}{\bf r}\cdot,
2i\bfnabla_r\right\} S\,.
\eea
Switching off the electromagnetic interaction, this expression should match 
Eq. \eqref{Lsinglet}, which fixes $\delta V_a =1/4$ and  $\delta V_b =0$ to all orders in perturbation theory. 
The fact that the kinetic energy in pNRQCD is protected against quantum corrections 
is a direct consequence of Poincar\'e invariance \cite{hep-ph/0306107}. We conclude, therefore, that  
\begin{equation}
V^{\nabla \cdot (r \times B)} = V^{\nabla _r \cdot (r \times (r\nabla)B)}  = 1 
\end{equation}
holds to all orders in perturbation theory.

\begin{figure}
  \centering
  \epsfig{file=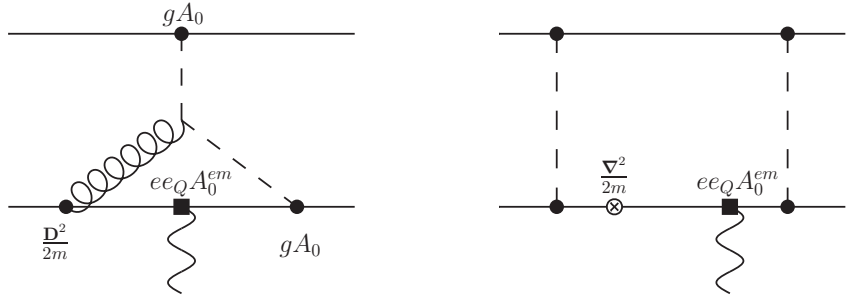,width=0.7 \linewidth,clip=}
  \caption{Examples of diagrams at order $1/m$ with an external longitudinal photon.}
  \label{Derivative_diagram2}
\end{figure}

3. The photon is longitudinal and coupled to the heavy quark/antiquark lines.
Such diagrams match only operators with electric fields.
If we consider diagrams with one insertion of a chromomagnetic dipole operator
or a kinetic energy operator (for the latter see  Fig.~\ref{Derivative_diagram2}), 
then such diagrams may possibly contribute at order $1/m$ to operators that vanish at
${\cal O}(\als^0)$, for instance, 
\be
\int d^3 r \; \frac{1}{mr} S^{\dagger} {\bf r}\cdot e e_Q {\bf E}^{\em} S \,.
\ee
In both cases, the argument developed in Sec.~\ref{secreducible}, paragraph 
{\it{(vi)}}, applies, for $A_0^{\em}$ commutes with gluons.
Hence, all such diagrams cancel in the matching against reducible 
pNRQCD diagrams, and an operator like the one written above does not show 
up even at higher orders in the coupling constant.

\subsection{Matching at $\mathcal{O} (1/m^2)$}
Diagrams that contribute to the matching at order $1/m^2$ 
contain, besides an arbitrary number of operators of order $1/m^0$, 
either two operators of order $1/m$ or one of the following 
operators of order $1/m^2$:
\begin{eqnarray*}
\frac{c_S}{8m^2} \boldsymbol \sigma \cdot [i{\bf D} \times, g{\bf E}], \quad 
\frac{c_S^{\em}}{8m^2} \boldsymbol \sigma \cdot [i{\bf D} \times, ee_Q{\bf E}^{\em}], \quad
\frac{c_D}{8m^2} [{\bf D} \cdot, g{\bf E}], \quad
\frac{c_D^{\em}}{8m^2} ee_Q (\bfnabla\cdot{\bf E}^{\em}) \,, 
\end{eqnarray*}
which we call the (chromo)electric spin-orbit operator and 
the (chromo)electric Darwin operator respectively.
Since many different diagrams can contribute, it is convenient to 
distinguish between spin-dependent and spin-independent amplitudes.
It is always implicitly assumed that diagrams may contain 
an arbitrary number of $1/m^0$ gluonic vertices.

\subsubsection{Spin-dependent diagrams} 
First, we consider the matching of spin-dependent operators. 

1. We consider diagrams made of two chromomagnetic dipole operators 
and a longitudinal photon. Since the photon commutes  
with the chromomagnetic dipole operators, the result of Sec.~\ref{secreducible}, paragraph 
{\it{(v)}}, applies: these diagrams do not contribute to the matching of new operators.

2. Diagrams that contain one magnetic dipole operator and a 
chromomagnetic dipole one are of the type discussed in Sec.~\ref{secreducible}, paragraph 
{\it{(vi)}}. Hence they do not contribute to the matching of new operators.
Moreover, in \cite{hep-ph/0512369}, it has been pointed out that this kind of amplitudes, 
proportional to the expectation value of a chromomagnetic field, vanishes for parity. 

3. The situation is similar for diagrams that contain one magnetic dipole operator and a kinetic energy one. 
Since the magnetic dipole operator commutes with the gluons, we are in the situation of 
Sec.~\ref{secreducible}, paragraph {\it{(vi)}}, and this type of diagrams 
does not contribute to the matching of new operators.

\begin{figure}
  \centering
  \epsfig{file=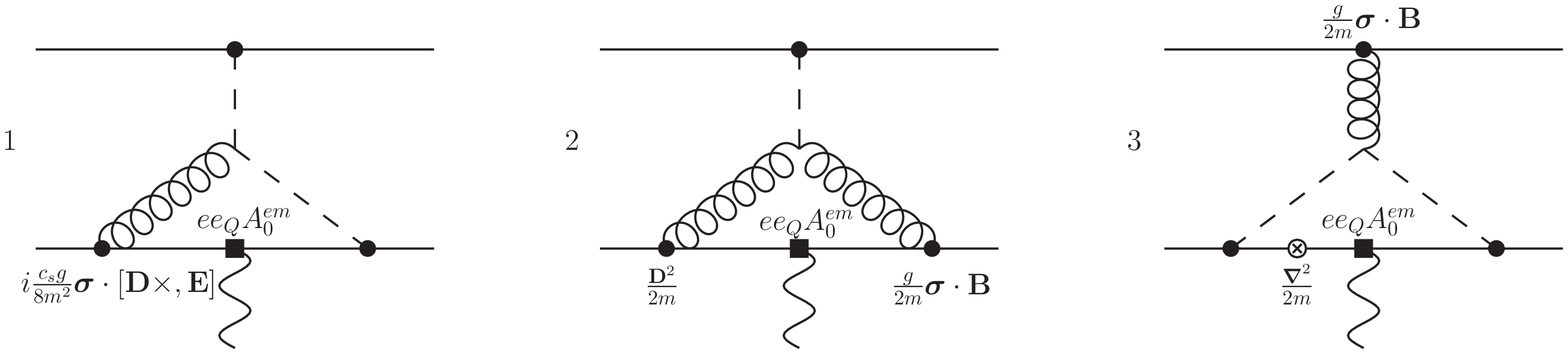,width= \linewidth,clip=}
  \epsfig{file=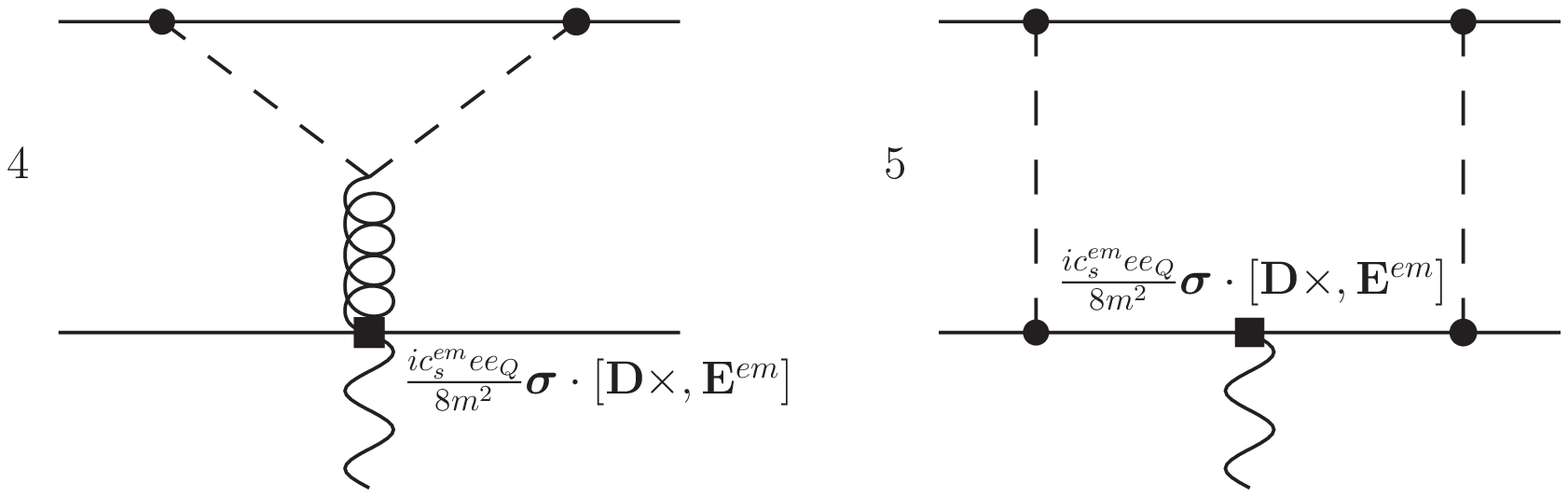,width=0.65 \linewidth,clip=}
  \caption{Some diagrams that contribute to spin-dependent operators of order $1/m^2$ containing an electric field.}
  \label{spin-dependent-operators}
\end{figure}

4. Also diagrams that contain  one chromoelectric spin-orbit operator and one longitudinal 
photon (see e.g. diagram 1 in Fig.~\ref{spin-dependent-operators}) fall 
under the situation discussed in Sec.~\ref{secreducible}, paragraph {\it{(vi)}}, 
and do not contribute to the matching of new operators.
 
5. We consider now diagrams made with one chromomagnetic dipole operator, 
one kinetic energy operator and a longitudinal photon 
(see e.g. diagrams 2 and 3 in Fig.~\ref{spin-dependent-operators}). 
Since the longitudinal photon commutes with the chromomagnetic dipole 
operator, the argument of Sec.~\ref{secreducible}, paragraph {\it{(vii)}},
applies and therefore such diagrams do not contribute to the matching 
of new operators in pNRQCD.

6. Diagrams that contain one electric spin-orbit operator 
(see e.g. diagrams 4 and 5 in Fig.~\ref{spin-dependent-operators}) 
contribute to operators that are at least $v^2$ suppressed 
with respect to the leading electric dipole operator of pNRQCD.
These operators depend on spin and on an electric field.
For time inversion invariance, they must also depend on the 
quarkonium momentum. Hence, only the derivative part of the covariant 
derivative of the electric spin-orbit operator contributes and 
only if it does not act on the gluon fields. Since 
$\displaystyle \frac{c_S^{\em}}{8m^2} \boldsymbol \sigma \cdot [-i{\bfnabla} \times, ee_Q{\bf E}^{\em}]$
effectively acts as an operator that commutes with the gluons, 
we are in the situation discussed 
in Sec.~\ref{secreducible}, paragraph {\it{(iv)}}, that led to 
the matching condition \eqref{coem}. In our case and at leading order in the multipole 
expansion, the matching condition becomes 
\be
V^{\sigma \cdot(E \times \nabla _r)} = c_S^{\em} \, ,
\ee
which is exact. Operators that come from higher-orders in the 
multipole expansion do not need to be considered here because they are beyond our accuracy
(although the matching fixes also the Wilson coefficients of those operators to all orders).
Operators that are possible for symmetry arguments alone, like for instance
\begin{equation*}
\frac{1}{m^2r^2} [ \mathrm{S}^{\dagger}, \bfsigma^i] 
\left\{({\bf r}\times i\bfnabla_r)^i, {\bf r}\cdot e e_Q{\bf E}^{\em} \right\}
\mathrm{S},  
\end{equation*}
cannot be generated at any order in the strong-coupling constant 
and, therefore, may be set to zero in pNRQCD.
Finally, we observe that the observation made in \cite{hep-ph/0002250}
that $\bra{0}{\bf D}_{x_1}\ket{0} = \bfnabla_{x_1}$ and $\bra{0}{\bf D}_{x_2}\ket{0} = \bfnabla_{x_2}$
would lead to the same result.

7. Diagrams containing either one chromomagnetic operator
and a kinetic energy operator coupled to an external photon or 
one chromoelectric spin-orbit operator with the electromagnetic field encoded in the
covariant derivative have been calculated to all orders in  \cite{hep-ph/0512369}
with an argument similar to the one used in Sec.~\ref{matching-1/m}, paragraph 2. 
They contribute to the Wilson coefficient of one single operator,
\be
\frac{1}{4 m^2}
\; \frac{1}{r^3} 
\; \left\{{\rm S}^\dagger , \bfsigma\cdot\left[ {\bf r} \times  
\left({\bf r}\times e e_Q {\bf B}^{\em} \right) \right] \right\} {\rm S}\,, 
\ee
which, however, is relevant only for M1 transitions and not for E1 ones, since 
it does not change the parity of the quarkonium state.

\subsubsection{Spin-independent diagrams}
Now we consider the matching of spin-independent operators. 

1. Diagrams with one insertion of an electric Darwin operator do 
not contribute beyond ${\cal O}(\als^0)$ to the matching, for the electric Darwin operator commutes 
with gluons and the conclusion of Sec.~\ref{secreducible}, paragraph {\it{(iv)}}, 
applies. These diagrams match into the electric Darwin operator of pNRQCD. 
Such an operator has not been displayed in Eq. \eqref{LagE1}, because 
it does not contribute to E1 transitions.

2. Diagrams that contain one chromoelectric Darwin operator and a longitudinal 
photon are of the type discussed in  Sec.~\ref{secreducible}, paragraph {\it{(v)}}.
The electromagnetic interaction factorizes and the contribution cancels in the matching.

3. We consider diagrams containing two kinetic energy operators.
The (transverse) electromagnetic field is embedded in one of the covariant derivatives.
These diagrams match spin-independent operators of pNRQCD with one magnetic field.
Because of the pNRQCD symmetry under charge conjugation and ${\bf x}_1 \leftrightarrow {\bf x}_2$ exchange, 
the allowed spin-independent operators with one magnetic field must contain an odd number of 
${\bf r}$ or $\boldsymbol \nabla_r$. Moreover, because of parity invariance, 
at least one center of mass derivative has to be present too. 
However, such operators are too much suppressed to be relevant for E1 transitions at relative order $v^2$, 
an example being the operator
\begin{equation*}
\frac{1}{m^2 r} {\rm S}^{\dagger} \{ i \boldsymbol \nabla \cdot , {\bf r} \times ee_Q{\bf B}^{\em} \} {\rm S} \,, 
\end{equation*}
which is of relative order $v^3$ with respect to the leading electric dipole operator.

\begin{figure}[t]
\centering
\epsfig{file=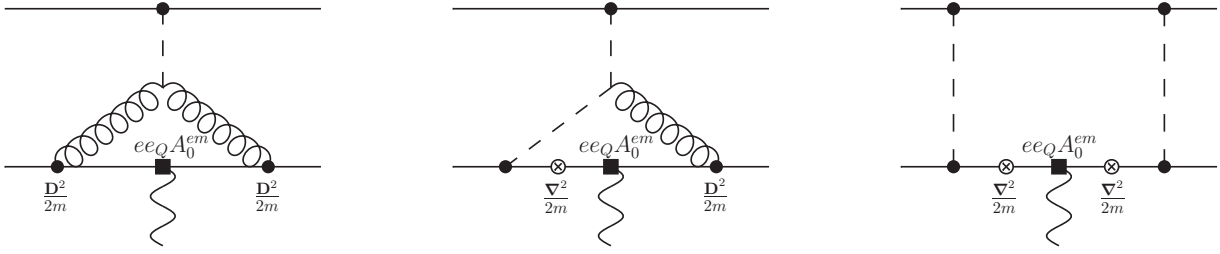,width=\linewidth,clip=}
\caption{Some diagrams contributing to the matching of spin-independent operators of order $1/m^2$.}
\label{spin-independent-operators}
\end{figure}

4. Finally, we consider diagrams containing two kinetic energy operators and an 
external longitudinal photon, see Fig.~\ref{spin-independent-operators}. 
We observe that $A_0^{\em}({\bf x}_1) + A_0^{\em\,c}({\bf x}_2) = 
{\bf r} \cdot \bfnabla A_0^{\em} + {\cal O}(r^2)$. 
Hence, these diagrams contribute to operators that are at least $1/m^2 \times (mv)^2 \sim v^2$ 
suppressed with respect to the leading electric dipole operator of pNRQCD; 
operators generated by higher orders in the multipole expansion are beyond our accuracy.
Let us consider now the kinetic energy operators.
If one of the derivatives acts on the longitudinal photon 
then the diagrams contribute to operators that are at least $v^3$ 
suppressed with respect to the leading electric dipole operator of pNRQCD.
Operators of relative order $v^3$ are beyond our accuracy.
If one of the derivatives acts on the factor ${\bf r}$ in ${\bf r} \cdot \bfnabla A_0^{\em}$ 
then the other derivatives do not. As a consequence, 
the electromagnetic operator ${\bf r} \cdot \bfnabla A_0^{\em}$ commutes at relative order $v^2$ 
with at least one of the kinetic energy operators and we are in the situation 
described in Sec.~\ref{secreducible}, paragraph {\it{(vii)}}. 
We conclude that in pNRQCD these diagrams cancel against iterations of lower-order 
potentials and electromagnetic vertices and do not contribute to any new operator.
In particular, possible operators allowed by the symmetries of pNRQCD, like, 
for instance,
\begin{equation*}
\frac{1}{m^2 r^2}  {\rm S}^{\dagger} {\bf r}\cdot e e_Q  {\bf E}^{\em} {\rm S} \, , 
\qquad
\frac{1}{m^2}  {\rm S}^{\dagger} \left\{\bfnabla_r^2,{\bf r}\cdot e e_Q  {\bf E}^{\em} \right\}{\rm S} \, ,
\end{equation*}
cannot be generated at any order in the strong-coupling constant.

\subsection{Concluding remarks}
In this section, we have matched non-perturbatively all operators of
pNRQCD relevant to describe E1 transitions at relative order $v^2$. It
turns out that the situation for E1 transitions is different from the
one for M1 transitions first discussed in \cite{hep-ph/0512369}.

For E1 transitions no $\mathcal{O}(1/m^3)$ operator is relevant at
relative order $v^2$, as it is shown in Eq. \eqref{LagE1}. 
In particular, we can neglect the matching of NRQCD diagrams with three
kinetic energy operator insertions or with one insertion of the
operator ${\bf D}^4/(8m^3)$.

For M1 transitions between strongly-coupled states, corrections of relative order $v^2$ 
require the non-perturbative matching of $1/m^3$ operators. At this order 
one can perform an exact matching for all but one relevant operator 
(see appendix). This is the operator 
\begin{equation} 
\label{m3operator} 
\frac{1}{m^3r^2} \, V^{\frac{\sigma\cdot B}{m^3}} \, \left\{ {\rm S}^\dagger , \bfsigma
\cdot e e_Q {\bf B}^{\em} \right\} {\rm S}\,,  
\end{equation} 
whose Wilson coefficient, $ V^{\frac{\sigma\cdot B}{m^3}}$, can possibly get QCD corrections 
at the momentum-transfer scale. These are encoded in the expectation value of some suitable
Wilson loop, whose explicit expression is at present unknown.

\section{E1 transitions}
\label{secDecay}
After having derived the relevant pNRQCD Lagrangian, we will now 
proceed in the calculation of the electric dipole transition rates, starting from the 
non-relativistic limit. This will be used to fix the notation and 
discuss the wave functions. We will follow closely~\cite{hep-ph/0512369}.

\subsection{Quarkonium states}
We consider radiative transitions, $H \rightarrow H'\,\gamma$, 
between a quarkonium $H$ and a quarkonium $H'$.
A quarkonium state $\lvert H ({\bf P},\lambda) \rangle $ is an eigenstate 
of the pNRQCD Hamiltonian with the quantum numbers of a quarkonium $H$ 
with polarization $\lambda$. We normalize it in the non-relativistic way, i.e.
\begin{equation}
\langle H({\bf P}',\lambda ')\lvert H ({\bf P},\lambda) \rangle = 
(2\pi)^3 \delta ^3 ({\bf P} - {\bf P}') \delta _{\lambda \lambda'} \, .
\label{H-normalization}
\end{equation}
The leading-order quarkonium state is defined as 
\begin{align}\label{state}
\lvert H ({\bf P},\lambda) \rangle^{(0)} = & \int d^3 R \int d^3 r \, e^{i \bf P \cdot R} 
\textrm{Tr} \left\{ \phi ^{(0)}_{H(\lambda)}({\bf r}) \textrm{S}^{\dagger} ({\bf r}, {\bf R}) \ket{\textrm{vac}} \right\} \, ,
\end{align}
which is also an eigenstate of the total spin, the orbital angular momentum 
and the center of mass momentum ${\bf P}$ of the quark-antiquark pair. 
The wave function $\phi^{(0)}_{H(\lambda)}({\bf r})$ 
is an eigenfunction of the leading pNRQCD singlet Hamiltonian $h^{(0)}_S$, 
i.e. a solution of the Schr\"odinger equation
\begin{equation}
h^{(0)}_S \phi ^{(0)}_{H(\lambda)} 
\equiv \left(-\frac{{\boldsymbol \nabla _r}^2}{m} + V_S^{(0)}\right)\phi ^{(0)} _{H(\lambda)} 
= E ^{(0)} _{H}  \phi ^{(0)} _{H(\lambda)} \, .
\end{equation}
The eigenvalue $E^{(0)}_{H}$ is the leading-order binding energy. 

The wave functions have the following angular structures for $L=0$ \cite{hep-ph/0512369}
\begin{eqnarray}
\label{wave1}
\phi ^{(0)} _{n^1 S_0 }({\bf r}) & = & \sqrt{\frac{1}{8 \pi}} \, R_{n0} (r) \, , \\
\phi ^{(0)} _{n^3 S_1 (\lambda)}({\bf r}) & = & \sqrt{\frac{1}{8 \pi}} \, R_{n0} (r) \, 
\boldsymbol \sigma \cdot {\bf e}_{n^3 S_1} (\lambda) \, ,
\end{eqnarray}
for $L=1$ 
\begin{eqnarray}
\phi ^{(0)} _{n^1 P_1 (\lambda)}({\bf r}) & = & 
\sqrt{\frac{3}{8 \pi}} \, R_{n1} (r)\, {\bf e}_{n^1 P_1} (\lambda) \cdot \hat{\bf r} \, , \\
\phi ^{(0)} _{n^3 P_0}({\bf r}) & = & 
\sqrt{\frac{1}{8 \pi}} \, R_{n1} (r) \boldsymbol \sigma \cdot \hat{\bf r} \, ,\\
\phi ^{(0)} _{n^3 P_1 (\lambda)}({\bf r}) & = & 
\sqrt{\frac{3}{16 \pi}} \, R_{n1} (r) \,  \boldsymbol \sigma \cdot (\hat{\bf  r} \times {\bf e}_{n^3 P_1} (\lambda)) \, ,\\
\phi ^{(0)} _{n^3 P_2 (\lambda)}({\bf r}) & = & 
\sqrt{\frac{3}{8 \pi}} \, R_{n1} (r) \, \boldsymbol \sigma ^i h^{ij}_{n^3 P_2} (\lambda) \hat{\bf r}^j \, ,
\end{eqnarray}
and for $L=2$ \cite{arXiv:1007.4541,Piotr}
\begin{eqnarray}
\phi ^{(0)} _{n^1 D_2 (\lambda)}({\bf r}) &=& 
\sqrt{\frac{15}{16 \pi}} \, R_{n2} (r) \, \hat{\bf r}^i h^{ij}_{n^1 D_2} (\lambda) \hat{\bf r}^j \, ,\\
\phi ^{(0)} _{n^3 D_1 (\lambda)}({\bf r}) &=& 
\sqrt{\frac{1}{16 \pi}} \, R_{n2} (r) \,  
\left[3 ({\bf e} _{n^3 D_1} (\lambda) \cdot \hat{\bf r}) (\boldsymbol \sigma \cdot \hat{\bf r}) 
- \boldsymbol \sigma \cdot {\bf e}_{n^3 D_1} (\lambda)\right] \, ,\\
\phi ^{(0)} _{n^3 D_2 (\lambda)}({\bf r}) &=& 
\sqrt{\frac{5}{8 \pi}} \, R_{n2} (r) \,  \hat {\bf r}^i h^{ij}_{n^3 D_2} (\lambda) ({\hat{\bf r} \times \boldsymbol \sigma})^j \, ,\\
\phi ^{(0)} _{n^3 D_3 (\lambda)}({\bf r}) &=& 
\sqrt{\frac{15}{16 \pi}} \, R_{n2} (r) \, \hat{\bf r}^i \hat{\bf r}^j H^{ijk}_{n^3 D_3} (\lambda) \boldsymbol \sigma ^k \, . 
\end{eqnarray}
The vectors ${\bf  e}_{n^{2S+1} L_J}(\lambda)$ denote orthonormal polarization vectors of the quarkonium state. 
The tensors $h^{ij}_{n^{2S+1} L_J}(\lambda)$ and $H^{ijk}_{n^{2S+1} L_J}(\lambda)$ are completely symmetric, traceless
(the tensor $H^{ijk}$ has vanishing partial traces, i.e. $ H^{iik}_{n^{2S+1} L_J}=0$) and normalized as
\begin{equation}
h^{ij\,\ast}_{n^{2S+1} L_J}(\lambda) h^{ij}_{n^{2S+1} L_J}(\lambda ')  
=  \delta _{\lambda \lambda '} = H^{ijk\,\ast}_{n^{2S+1} L_J}(\lambda) H^{ijk}_{n^{2S+1} L_J}(\lambda ') \, .
\end{equation}

Whereas the quarkonium state is normalized in the non-relativistic fashion 
\eqref{H-normalization}, the one photon state, $ \lvert\gamma ({\bf k}, \sigma) \rangle$,
is normalized in the usual Lorentz-invariant way 
\begin{equation}
\langle\gamma ({\bf k}, \sigma) \lvert\gamma ({\bf k}', \sigma ') \rangle =
2k (2\pi)^3 \delta ^3({\bf k}-{\bf k}') \delta _{\sigma \sigma '} \, .
\label{photon-normalization}
\end{equation}
This implies that external electric or magnetic fields project on a one photon state as 
\bea
\langle\gamma ({\bf k}, \sigma) \lvert {\bf E}^{\em}({\bf R}) \lvert \textrm{vac}\rangle &=& 
-ik{\boldsymbol \epsilon}^{\ast} (\sigma) e^{-i{\bf k} \cdot {\bf R}} \,,
\label{Epro}\\
\langle \gamma({\bf k},\sigma) | {\bf B}^{\em}({\bf R}) | \textrm{vac}\rangle  &=& 
-i \, {\bf k} \times \bfepsilon^*(\sigma)\,e^{-i{\bf k}\cdot{\bf R}} \,,
\label{Bpro}
\eea
where ${\boldsymbol \epsilon}(\sigma)$ is the photon polarization vector 
(the dependence of ${\boldsymbol \epsilon}(\sigma)$ on ${\bf k}$ is understood).
The photon transversality requires 
${\boldsymbol \epsilon}(\sigma) \cdot {\bf k} =0$. 

Finally, the quarkonium and photon polarizations satisfy the relations 
\begin{eqnarray}
\sum _{\lambda} {\bf  e}^{i\,\ast}_{n^{2S+1} L_J}(\lambda) {\bf  e}^j_{n^{2S+1} L_J}(\lambda) &=& \delta ^{i j} \,, 
\label{pol1} \\
\sum _{\lambda} h^{ij\,\ast}_{n^{2S+1} L_J}(\lambda) h^{kl}_{n^{2S+1} L_J}(\lambda) &=& 
\frac{1}{2}(\delta ^{ik} \delta ^{jl} + \delta ^{il} \delta ^{jk}) -\frac{1}{3} \delta ^{ij} \delta ^{kl} \,,
\\
\sum _{\lambda} H^{ijk\,\ast}_{n^{2S+1} L_J}(\lambda) H^{lmn}_{n^{2S+1} L_J}(\lambda) &=& 
\frac{1}{(3 !)^2} \left(\delta ^{il}\delta ^{jm}\delta ^{kn} -\frac{3}{5}\delta ^{ij}\delta ^{kl}\delta ^{mn} \right.
\nonumber \\
& &  + \,\hbox{permutations of the indices $i,j,k$ and $l,m,n$}
\Big) \, ,
\\ 
\sum _{\sigma} {\boldsymbol \epsilon}^i(\sigma)   {\boldsymbol \epsilon}^{j\,\ast }(\sigma)  &=& 
\delta^{ij} - \hat{\bf k}^{i}\hat{\bf k}^{j}\,. 
\label{pol4} 
\end{eqnarray}

\begin{figure}
  \centering
  \epsfig{file=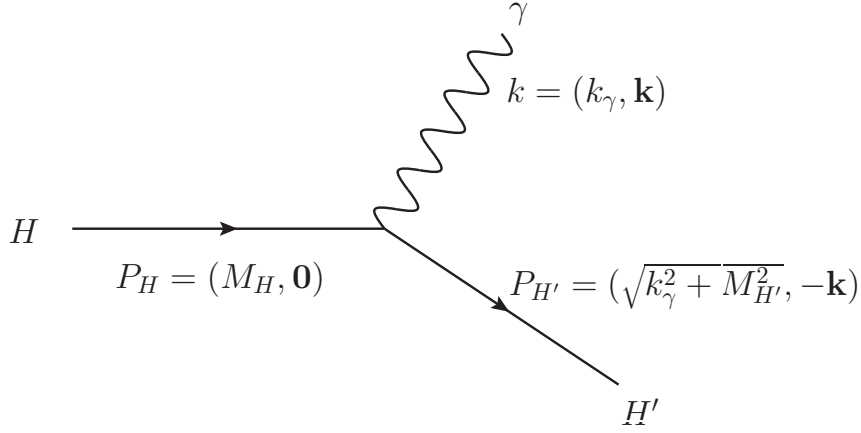,width=0.7\linewidth,clip=}
  \caption{Kinematics of the transition $H \rightarrow H'\,\gamma$ in the center of mass frame.}
  \label{kinematics}
\end{figure}

\subsection{Transition amplitudes and rates}
In the rest frame of the initial quarkonium, see Fig.~\ref{kinematics}, 
the transition amplitude from a quarkonium $H$ with polarization $\lambda$ 
to a quarkonium $H'$ with momentum ${\bf P}'$ and polarization $\lambda '$, and a photon with energy
\begin{equation}
k_{\gamma} = \lvert {\bf k} \rvert =  \frac{M_H^2 -M_{H'}^2}{2M_H} = (M_H - M_H') + \, \mathcal{O}\left(\frac{k_{\gamma}^2}{M_H} \right), 
\end{equation}
and polarization $\sigma$ is given by 
\bea
{\cal A}\left[H({\bf 0},\lambda) \to H^\prime(-{\bf k},\lambda^\prime) \gamma({\bf k},\sigma)\right] 
(2\pi)^3 \delta ^3 ({\bf P} ' + {\bf k}) &=& 
\nonumber\\
&& \hspace{-4cm}
- \langle H'({\bf P}',\lambda ')  \gamma ({\bf k}, \sigma) \lvert 
\int d^3 R \, \mathcal{L}_{\gamma \textrm{pNRQCD}} \, \lvert H ({\bf 0},\lambda) \rangle \, .
\label{amplitude}
\eea
The corresponding transition width reads
\begin{align}
\label{decay rate}
\Gamma_{H \rightarrow H' \,\gamma} = 
& \, \int \frac{d^3 P'}{(2 \pi)^3} \frac{d^3 k}{(2 \pi)^3} \frac{1}{2k} \, (2 \pi)^4 \delta ^4(P_H - k -P') \, 
\frac{1}{N_{\lambda}} \sum _{\lambda \lambda ' \sigma} \lvert \mathcal{A}_{H \rightarrow H' \,\gamma} \rvert^2 
\nonumber \\
= & \, \frac{1}{8\pi ^2} \left( 1- \frac{k_{\gamma}}{M_H}\right) 
\int_0^\infty \!\! dk \, k \int d \Omega({\bf \hat{k}}) \, \delta (k- k_{\gamma}) 
\frac{1}{N_{\lambda}} \sum _{\lambda \lambda ' \sigma} \lvert \mathcal{A}_{H \rightarrow H' \,\gamma} \rvert ^2 \, ,
\end{align}
where the initial state is averaged over the polarizations, whose number is $N_\lambda$, and 
$ \mathcal{A}_{H \rightarrow H' \,\gamma}$ is a short-hand notation for 
${\cal A}\left[H({\bf 0},\lambda) \to H^\prime(-{\bf k},\lambda^\prime) \gamma({\bf k},\sigma)\right]$.

\subsection{The non-relativistic limit}
The leading operator responsible for E1 transitions in pNRQCD is the electric dipole operator 
\begin{equation}
\mathcal{L}_{E1} = \int d^3 r \, \mathrm{Tr} \, \left\{ {\rm S}^{\dagger} {\bf r}\cdot  e e_Q{\bf E}^{\em} {\rm S} \right\} \, .
\label{leadingE1}
\end{equation}
From Eqs. \eqref{state}, \eqref{Epro}, \eqref{amplitude} and the equal-time commutation rules for singlet fields 
(spin indices are written down explicitly)
\begin{equation}
\left[{S}^{\dagger}_{ij}({\bf r}, {\bf R}, t), {S}_{kl}({\bf r}', {\bf R}', t)\right]= 
\delta _{il}\delta _{jk} \delta^3 ({\bf r} - {\bf r}') \delta^3 ({\bf R}-{\bf R}'),
\end{equation}
it follows that the leading amplitudes for the E1 transitions 
$n^3P_J \rightarrow n^{\prime\,3}S_1 \,\gamma$ and $n^1P_1 \rightarrow n^{\prime\,1}S_0 \,\gamma$ read 
\begin{align}
& \mathcal{A}^{(0)}_{n^3P_0 \rightarrow n^{\prime\,3}S_1 \,\gamma}= \frac{i e e_Q k}{3} \, I_3(n1 \rightarrow n'0) \, 
{\boldsymbol \epsilon}^{\ast}(\sigma) \cdot {\bf e}_{n^{\prime\,3} S_1}  (\lambda ') \, ,
\\
& \mathcal{A}^{(0)}_{n^3P_1 \rightarrow n^{\prime\,3}S_1 \,\gamma} = \frac{i e e_Q k}{\sqrt{6}} \, I_3(n1 \rightarrow n'0) \, 
{\boldsymbol  \epsilon}^{\ast}(\sigma) \cdot ({\bf e}_{n^3 P_1} (\lambda) \times {\bf e}_{n^{\prime\,3} S_1} (\lambda ')) \, , 
\\
& \mathcal{A}^{(0)}_{n^3P_2 \rightarrow n^{\prime\,3}S_1 \,\gamma} = \frac{i e e_Q k}{\sqrt{3}} \,  I_3(n1 \rightarrow n'0) \, 
{\bf e}_{n^{\prime\,3} S_1}^{i}(\lambda ') h^{ij}_{n^3 P_2} (\lambda){\boldsymbol \epsilon}^{\ast \, j}(\sigma) \,, 
\\
& \mathcal{A}^{(0)}_{n^1P_1 \rightarrow n^{\prime\,1}S_0 \,\gamma}= \frac{i e e_Q k}{\sqrt{3}} \, I_3(n1 \rightarrow n'0) \, 
{\boldsymbol \epsilon}^{\ast}(\sigma) \cdot {\bf e}_{n^1 P_1}  (\lambda) \,,
\end{align}
where we have defined 
\bea
I_N (nL \rightarrow n'L')&=& \int_0^{\infty} dr \, r^N \, R_{n'L'}(r) \, R_{nL}(r) \, ,
\\
I^\prime_N (nL \rightarrow n'L')&=& \int_0^{\infty} dr \, r^N \, R_{n'L'}(r) \, \frac{d}{d r}R_{nL}(r) \,.
\eea
The second definition is for further use. We also give the amplitudes for $n^3 D_1 \rightarrow n^{\prime\,3}P_J \,\gamma$:
\begin{align}
\label{D-transition1}
& \mathcal{A}^{(0)}_{n^3D_1 \rightarrow n^{\prime\,3}P_0 \,\gamma}  = 
\frac{i e e_Q k \sqrt{2}}{3} \, I_3(n2 \rightarrow n'1) 
\, {\boldsymbol  \epsilon}^{\ast}(\sigma) \cdot {\bf e}_{n^3 D_1} (\lambda) \, , 
\\
& \mathcal{A}^{(0)}_{n^3D_1 \rightarrow n^{\prime\,3}P_1 \,\gamma} = 
\frac{i e e_Q k}{2\sqrt{3}} \, I_3(n2 \rightarrow n'1) 
\, {\boldsymbol  \epsilon}^{\ast}(\sigma) \cdot ({\bf e}_{n^3 D_1} (\lambda) \times {\bf e}_{n^{\prime\,3} P_1} (\lambda ')) \, , 
\\
& \mathcal{A}^{(0)}_{n^3D_1 \rightarrow n^{\prime\,3}P_2 \,\gamma} = 
\frac{i e e_Q k}{5\sqrt{6}} \, I_3(n2 \rightarrow n'1) \, 
{\bf e}_{n^{3} D_1}^{i}(\lambda) h^{ij}_{n^{\prime\,3} P_2} (\lambda'){\boldsymbol \epsilon}^{\ast \, j}(\sigma) 
\label{D-transition3} \,.
\end{align}

From (\ref{decay rate}) and the relations (\ref{pol1}-\ref{pol4}), it follows that the non-relativistic decay rates are
\bea
\label{E1_0}
\Gamma^{(0)}_{n^3 P _J \rightarrow n^{\prime\,3} S_1 \,\gamma} &=& \Gamma^{(0)}_{nn'} \left( 1- \frac{k_{\gamma}}{M_{n^3 P_J}} \right)\,,
\\
\Gamma^{(0)}_{n^1 P _{1} \rightarrow n^{\prime\,1} S_0 \,\gamma} &=& \Gamma^{(0)}_{nn'} \left( 1- \frac{k_{\gamma}}{M_{n^1 P_1}} \right)\,,
\eea
with
\begin{equation}
\Gamma^{(0)}_{nn'} \equiv 
\frac{4}{9} \, \alpha e_Q ^2 k_{\gamma}^3 [I_3(n1 \rightarrow n'0)]^2 \,  \sim \frac{k_{\gamma}^3}{m^2 v^2} \, .
\end{equation}
E1 transition rates are of order ${k_{\gamma}^3}/{(m^2 v^2)}$, which means that they happen $1/v^2$ more 
frequently than allowed M1 transitions at the same photon energy. 
The terms proportional to $k_{\gamma}/M_{n^3 P_J} \sim k_{\gamma}/M_{n^1 P_1} \sim mv^2/(2m)$ 
are suppressed by $v^2$ and can therefore be neglected at leading order.
They contribute and have to be accounted for at relative order $v^2$, which we will compute in the next section.
We note that at leading order the transition rate is independent of the initial total angular momentum $J$.
The reverse transitions depend instead on the final total angular momentum: 
\begin{align}
\Gamma^{(0)}_{n^3 S_1 \rightarrow n^{\prime\,3} P_J \,\gamma}= & \, \frac{2J+1}{3} \Gamma^{(0)}_{nn'} \left( 1- \frac{k_{\gamma}}{M_{n^3 S_1}} \right)
\,,\\
\Gamma^{(0)}_{n^1 S_0 \rightarrow n^{\prime\,1} P_1 \,\gamma}= & \, 3 \Gamma^{(0)}_{nn'} \left( 1- \frac{k_{\gamma}}{M_{n^1 S_0}} \right)\, .
\end{align}
These results agree, neglecting $k_\gamma/m$ corrections, with the general non-relativistic formula~\cite{Eichten:1978tg}
\begin{equation}
\Gamma^{(0)}_{n^{2S+1} L _{J} \rightarrow n'^{2S+1} L' _{J'} \,\gamma}= 
\frac{4}{3} \, \alpha e_Q^2\, k_{\gamma}^3\, [I_3(nL \rightarrow n'L')]^2\, 
(2J'+1) \, \textrm{max}(L,L') \,  
\left\{
\begin{array}{ccc}
J & 1 & J' \\
L' & S & L
\end{array}
\right\}^2  ,
\end{equation}
where the last term in the brackets denotes a Wigner 6-j coefficient.

\section{Relativistic corrections to E1 transitions}
\label{secRel}
One of the main advantages of using pNRQCD is that it accounts for the corrections 
to the decay amplitudes in a systematic fashion. We will split the analysis into a part
that accounts for the electromagnetic interaction terms in the pNRQCD 
Lagrangian suppressed by $\mathcal{O}(v^2)$ with respect to the leading 
electric dipole operator \eqref{leadingE1}, and a part that accounts 
for $v^2$ corrections to the quarkonium state.
We will concentrate on the decay $n^3 P_J \rightarrow n^{\prime \, 3} S_1 \,\gamma$ 
to compare with the result in \cite{PRINT-84-0308}. 
The extension to other processes like 
$n^1 P_1 \rightarrow n^{\prime\,1} S_0 \,\gamma$, 
$n^3 S_1 \rightarrow n^{\prime \, 3} P_J \,\gamma$ 
and $n^1 S_0 \rightarrow n^{\prime \, 1} P_1 \,\gamma$ is straightforward.  
Final results for all these radiative transitions will be given in Sec.~\ref{secRes}.
For a non-vanishing leading-order transition amplitude
$\mathcal{A}^{(0)}_{n^{2S+1} L_J (\lambda) \rightarrow n^{\prime \, 2S+1}{L'}_{J'} (\lambda') \,\gamma}$,  
we define 
\begin{equation}
\bar{\mathcal{A}} \equiv  
\frac{\mathcal{A}}{\mathcal{A}^{(0)}} \,, \qquad \bar{{\Gamma}} \equiv \frac{\Gamma}{\Gamma^{(0)}} \, . 
\end{equation}

\subsection{Corrections induced by E1 operators of  relative order $v^2$}
At subleading order in the decay rate, i.e. at order
$k_{\gamma}^3/m^2$, all the interaction terms displayed in the
Lagrangian (\ref{LagE1}) beyond the leading electric dipole operator contribute.

\begin{itemize}
\item[(1)]{
The correction induced by the operator 
$S^{\dagger} {\bf r}\cdot [({\bf r}\cdot{ \boldsymbol \nabla})^2  e e_Q  {\bf E}^{\em}] S/24$ is 
\begin{equation}
\bar{\mathcal{A}}^{(1)}_{n^3P_J \rightarrow n^{\prime\,3}S_1 \,\gamma}= 
- \frac {k^2}{120} \frac {I_5(n1 \rightarrow n'0)}{I_3(n1 \rightarrow n'0)} \, .
\end{equation}
}
\item[(2)]{
The correction induced by the operator 
$i S^{\dagger} \{ {\boldsymbol \nabla} \cdot , {\bf r} \times e e_Q  {\bf B}^{\em} \} S/(4m)$ is
\begin{equation}
\bar{\mathcal{A}}^{\textrm{(2)}}_{n^3P_J\rightarrow n^{\prime\,3}S_1 \,\gamma}= \frac{k}{4m} \, .
\end{equation}
}
\item[(3)]{
The operator 
$i S^{\dagger} \{ {\boldsymbol \nabla}_r \cdot , {\bf r} 
\times  [({{\bf r} \cdot {\boldsymbol \nabla}}) e e_Q  {\bf B}^{\em}] \} S/(12 m)$ 
corrects the leading-order amplitude by an amount
\begin{equation}
\bar{\mathcal{A}}^{\textrm{(3)}}_{n^3P_J\rightarrow n^{\prime\,3}S_1 \,\gamma}= -\frac{k}{12m} \, .
\end{equation}
}
\item[(4)]{
The spin-dependent operator 
$c_F^{\em}[S^{\dagger},{\boldsymbol \sigma}] \cdot [({\bf r} \cdot {\boldsymbol \nabla}) e e_Q {\bf B}^{\em}]S/(4m)$ 
contributes with
\begin{align}
& \bar{\mathcal{A}}^{(4)}_{n^3P_0 \rightarrow n^{\prime\,3}S_1 \,\gamma}= 
\frac{k c_F^{\em}}{2m} \, ,
\\
& \bar{\mathcal{A}}^{(4)}_{n^3P_1 \rightarrow n^{\prime\,3}S_1 \,\gamma}= 
\frac{k c_F^{\em}}{2m}\frac{(\hat {\bf k} \cdot  {\bf e}_{n^{\prime\,3} S_1})
{\boldsymbol \epsilon}^{\ast} \cdot (\hat{\bf k} \times 
 {\bf e}_{n^3 P_1})}{ {\boldsymbol \epsilon}^{\ast} \cdot ( {\bf e}_{n^{\prime\,3} S_1} \times {\bf e}_{n^3 P_1})} \, , 
\\
& \bar{\mathcal{A}}^{(4)}_{n^3P_2 \rightarrow n^{\prime\,3}S_1 \,\gamma}= 
\frac{k c_F^{\em}}{2m} \frac{\left[({\boldsymbol \epsilon}^{\ast} \cdot 
 {\bf e}_{n^{\prime\,3} S_1}) \hat {\bf k}^{i} - (\hat {\bf k} \cdot  {\bf e}_{n^{\prime\,3} S_1}) 
{\boldsymbol \epsilon}^{\ast \, i} \right] h^{ij}_{n^3 P_2}
\hat{\bf k}^{j}}{{\boldsymbol \epsilon}^{\ast \, i} h^{ij}_{n^3 P_2}  {\bf e}^{j}_{n^{\prime\,3} S_1}} \,,
\end{align}
where the quarkonium and photon polarization numbers are not explicitly shown. 
Using Eqs. \eqref{pol1}-\eqref{pol4} the corresponding decay rates yield
\begin{equation}
\bar{{\Gamma}}^{(4)}_{n^3P_J \rightarrow n^{\prime\,3}S_1 \,\gamma} = - \frac{k_{\gamma} c_F^{\em}}{2m}\left[\frac{J(J+1)}{2}-2\right] \, .
\end{equation}
}
\item[(5)]{
Finally, the contribution of the operator 
$-i c_S^{\em}[S^{\dagger}, {\boldsymbol \sigma} ] \cdot (e e_Q {\bf E}^{\em} \times {\boldsymbol \nabla}_{r}) S/(4m^2)$ is 
\be
\bar{\mathcal{A}}^{(5)}_{n^3P_J \rightarrow n^{\prime\,3}S_1 \,\gamma} = 
\frac{c_S^{\em}}{2m^2} \left[\frac{J(J+1)}{2} -2 \right] 
\frac{I_2 '(n1 \rightarrow n'0) + 2 I_1 (n1 \rightarrow n'0)}{I_3 (n1 \rightarrow n'0)} \, .
\ee
}
\end{itemize}

\subsection{Quarkonium state corrections of  relative order $v^2$}
\label{wave-function-corrections}
The quarkonium state \eqref{state} is not an eigenstate of the complete 
Hamiltonian of pNRQCD. The eigenstate may be constructed from \eqref{state}
by systematically adding higher-order corrections, which are perturbative in the 
relative velocity $v$. Corrections may come from higher-order potentials 
($1/m$ and $1/m^2$ terms) and from higher Fock states, which account for the coupling 
of the quark-antiquark singlet state to other low-energy degrees of freedom. 
We have to include such corrections at relative order $v^2$, both in the initial 
and in the final quarkonium states, in order to achieve a precision of relative order $v^2$ 
in the E1 transition rates.

\subsubsection{Corrections due to higher-order potentials}
\label{sechopot}
The first-order correction to the quarkonium state \eqref{state} 
induced by a correction $\delta h_S$ to the Hamiltonian $h_S^{(0)}$ is given by
\bea
\lvert H ({\bf P},\lambda) \rangle ^{(1)} &=& 
\int \frac{d^3P^\prime}{(2\pi)^3}\,\sum _{H' \neq H, \lambda '} 
\frac{|H' ({\bf P}^\prime,\lambda')\rangle^{(0)}\; {}^{(0)}\langle H'({\bf P}^\prime,\lambda')|}{E_H^{(0)}-E_{H^\prime}^{(0)}}   
\nonumber\\
&& \hspace{3cm}\times
\int d^3 R \int d^3r \, \textrm{Tr} \left\{\textrm{S}^{\dagger} \delta h_S(r) \textrm{S} \right\} |H ({\bf P},\lambda)\rangle^{(0)}
\,.
\eea
We assume that, in order to account for corrections of relative order $v^2$,
we need to include in $\delta h_S$ all the $1/m$ and $1/m^2$ potentials and,  
at order $1/m^3$, the first relativistic correction to the kinetic energy. 
Such a counting, which holds for weakly-coupled quarkonia, appears to be generally consistent 
with heavy quarkonium spectroscopy \cite{hep-ph/9407339}, and it is indeed the most widely used. 
It should be remarked, however, that in the case of strongly-coupled quarkonia
it is not the most conservative one  \cite{hep-ph/0009145}.

In the assumed power counting, at relative order $v^2$, $\delta h_S$ has the form
\begin{equation}
\delta h_S (r) = -\frac{{\bf p}^4}{4m^3} + \delta V_S(r)\,,
\end{equation}
where ${\bf p} = -i \boldsymbol \nabla _r$. $\delta V_S$ is organized in powers of $1/m$, 
\be
\delta V_S(r) = \frac{V^{(1)}(r)}{m} + \frac{V^{(2)}_{\rm SI}(r)}{m^2}+ \frac{V^{(2)}_{\rm SD}(r)}{m^2} \, ,
\label{m2pot}
\ee
where, at order $1/m^2$, we have distinguished between spin independent (SI) 
and spin dependent (SD) terms \cite{hep-ph/0410047}, 
\begin{align}
V^{(2)}_{\rm SI}(r) = 
& V^{(2)}_r(r) + \frac{1}{2}\{ V^{(2)}_{p^2}(r), {\bf p}^2 \} + \frac{V^{(2)}_{L^2}(r)}{r^2}{\bf L}^2 \, ,
\\
V^{(2)}_{\rm SD}(r) = 
& V^{(2)}_{LS}(r){\bf L}\cdot {\bf S} +V^{(2)}_{S^2}(r){\bf S}^2 + V^{(2)}_{S_{12}}(r){\bf S}_{12} (\hat{\bf r}) 
\label{SD_potentials} \, ,
\end{align}
with ${\bf S} = {\bf S}_1 + {\bf S}_2 = ({\boldsymbol \sigma}_1 + {\boldsymbol \sigma}_2)/2$, ${\bf L}={\bf r} \times {\bf p}$
and ${\bf S}_{12} (\hat{\bf r}) =3 (\hat{\bf r}\cdot \boldsymbol\sigma _1)(\hat{\bf r}
\cdot \boldsymbol\sigma _2) - \boldsymbol\sigma _1 \cdot \boldsymbol\sigma_2$. 
Terms involving the center of mass momentum, which is $-{\bf k}_\gamma$ for the 
quarkonium in the final state, are suppressed by an extra $v$ and have been neglected.
In the weak-coupling case, the above potentials read at leading (non-vanishing)
order in perturbation theory (see e.g. \cite{hep-ph/9910238})
\begin{eqnarray}
&& \hspace{-4mm} 
V^{(1)}(r) =  - \frac{C_F N_c\als^2}{2r^2} \,, \quad 
V^{(2)}_r(r) =  \pi C_F \als \delta ^{3}({\bf r}) \,, \quad 
V^{(2)}_{p^2}(r) =  - \frac{C_F \als}{r} \,, \quad 
V^{(2)}_{L^2}(r) =  \frac{C_F \als}{2r} \,,
\nonumber\\
&& \hspace{-4mm} 
V^{(2)}_{LS}(r) =  \frac{3C_F \als}{2r^3} \,, \quad 
V^{(2)}_{S^2}(r) =  \frac{4 \pi C_F \als}{3} \delta^{3}({\bf r}) \,, \quad 
V^{(2)}_{S_{12}}(r) =  \frac{C_F \als}{4r^3} \,.
\label{pertpot}
\end{eqnarray}
In the strong-coupling case, the potentials are non-perturbative 
and can be expressed in terms of Wilson loops to be eventually 
evaluated on the lattice~\cite{hep-ph/0009145}.

There are, however, some observations that can be made without relying 
on any specific form of the potential, but just on its general structure \eqref{m2pot}-\eqref{SD_potentials}.
Let us consider the radiative transition $n^3P_J \rightarrow n^{\prime\,3}S_1 \,\gamma$.
Initial state corrections due to $\delta h_S$ read
\bea
\mathcal{A}^{\textrm{i}}_{n^3P_J \rightarrow n^{\prime\,3}S_1 \,\gamma} 
&=& - {}^{(0)}\langle n^{\prime\,3}S_1 \,\gamma \lvert \int d^3 R \; \mathcal{L}_{E1} \, \lvert n^3P_J \rangle^{(1)} 
\nonumber\\
&=& 
\sum_{m\neq n} \frac{{}^{(0)}\langle m^3P_J \lvert \delta h_S (r) \lvert n^3P_J  \rangle^{(0)}}{E^{(0)}_{n1} - E^{(0)}_{m1}} 
\, \mathcal{A}^{(0)}_{m^3P_J \rightarrow n^{\prime\,3}S_1 \,\gamma}\,, 
\label{initialstate}
\eea
where $E^{(0)}_{nL}$ is the leading-order binding energy of a quarkonium with principal quantum 
number $n$ and orbital angular momentum $L$.  Final state corrections read
\bea
\mathcal{A}^{\textrm{f}}_{n^3P_J \rightarrow n^{\prime\,3}S_1 \,\gamma} 
&=& - {}^{(1)}\langle n^{\prime\,3}S_1 \gamma \lvert \int d^3 R \; \mathcal{L}_{E1} \, \lvert n^3P_J \rangle^{(0)} 
\nonumber\\
&=& 
\sum_{m\neq n'}\frac{{}^{(0)}\langle n^{\prime\,3}S_1 \lvert\delta h_S (r) \lvert m^3S_1\rangle^{(0)}}{E^{(0)}_{n'0} - E^{(0)}_{m0}} 
\, \mathcal{A}^{(0)}_{n^3P_J \rightarrow m^{3}S_1 \,\gamma} 
\nonumber\\
&&
+ \sum_{m,J'} 
\frac{{}^{(0)}\langle n^{\prime\,3}S_1 \lvert \delta h_S(r) \lvert m^3D_{J'} \rangle^{(0)}}{E^{(0)}_{n'0} - E^{(0)}_{m2}} \, 
\mathcal{A}^{(0)}_{n^3P_J \rightarrow m^{3}D_{J'}\,\gamma} \, .  
\eea
In both expressions, a sum over the intermediate state polarizations is understood. 
We have made use of the selection rule for the electric dipole matrix element 
and of the fact that the potentials \eqref{m2pot}-\eqref{SD_potentials} 
do not change the total angular momentum of the state, which follows from \cite{BS}
\bea
\langle n^{2S+1}L_J|\, V^{(2)}_{LS}(r) \, {\bf L}\cdot{\bf S}  \, |n^{\prime\, 2S'+1}L'_{J'} \rangle &=& 
\delta_{LL'}\delta_{JJ'}\delta_{SS'} \frac{J(J+1) - L(L+1) - S(S+1)}{2}
\nonumber\\
&& \hspace{2.7cm}\times \langle n^{2S+1}L_J|\, V^{(2)}_{LS}(r) \, |n^{\prime\,2S+1}L_J \rangle,
\\
\langle n^{2S+1}L_J|\, V^{(2)}_{S^2}(r) \, {\bf S}^2  \, |n^{\prime\, 2S'+1}L'_{J'} \rangle &=& 
\delta_{LL'}\delta_{JJ'}\delta_{SS'} \; S(S+1)\;
\langle n^{2S+1}L_J|\, V^{(2)}_{S^2}(r) \, |n^{\prime\,2S+1}L_J \rangle,
\nonumber\\
\\
\langle n^{2S+1}L_J|\, V^{(2)}_{S_{12}}(r) \, {\bf S}_{12}(\hat{\bf r})   \, |n^{\prime\, 2S'+1}L_{J'} \rangle &=& 
\delta_{JJ'}\delta_{SS'}\langle {\bf S}_{12} \rangle_{LJS} 
\langle n^{2S+1}L_J|\, V^{(2)}_{S_{12}}(r) \, |n^{\prime\,2S+1}L_J \rangle,
\eea
where $\langle {\bf S}_{12} \rangle_{101} = -4$, $\langle {\bf S}_{12} \rangle_{111} = 2$,
$\langle {\bf S}_{12} \rangle_{121} = -2/5$, $\langle {\bf S}_{12} \rangle_{0JS}=0$ and 
$\langle {\bf S}_{12} \rangle_{LJ0}=0$.  
For the final state correction, we have assumed a completely non-degenerate spectrum.
To complete the spin structure of the wave-function corrections, we need to compute 
${}^{(0)}\langle n^3S_1 \lvert \, V^{(2)}_{S_{12}}(r) \, \lvert m^3D_{J} \rangle^{(0)}$.
It turns out that this matrix element does not vanish only for $J=1$ and in this case
\be
{}^{(0)}\langle n^3S_1 \lvert \, V^{(2)}_{S_{12}}(r) \, \lvert m^3D_{1} \rangle^{(0)}
= - 2\sqrt{2} \, {\bf e}_{m^3D_{1}}\cdot{\bf e}_{n^3S_1} \int_0^\infty dr\,r^2\, R_{n0}(r)R_{m2}(r)\, V^{(2)}_{S_{12}}(r)\,.
\ee
Summing over the intermediate state polarizations allows to factorize the leading order 
amplitude. For instance, in the case of the contribution of the spin-tensor potential 
$V^{(2)}_{S_{12}}(r){\bf S}_{12}(\hat{\bf r}) /m^2$ to the initial and final states, we obtain 
 \begin{align}
\bar{\mathcal{A}}_{n^3P_J \rightarrow n^{\prime\,3}S_1 \,\gamma}^{\textrm{i}}\Big|_{S_{12}} = 
& \frac{1}{m^2} \sum_{m \neq n} 
\frac{\displaystyle \int_0^\infty dr\,r^2\, R_{m1}(r)R_{n1}(r)\, V^{(2)}_{S_{12}}(r)}{E^{(0)}_{n1} - E^{(0)}_{m1}} 
\frac{I_3 (m1 \rightarrow n'0)}{I_3 (n1 \rightarrow n'0)} 
\langle {\bf S}_{12} \rangle_{1J1}\,,
\label{ST1}\\
\bar{\mathcal{A}}_{n^3P_J \rightarrow n^{\prime\,3}S_1 \,\gamma}^{\textrm{f}}\Big|_{S_{12}} = 
& \frac{1}{m^2} \sum_{m} 
\frac{\displaystyle \int_0^\infty dr\,r^2\, R_{n'0}(r)R_{m2}(r)\, V^{(2)}_{S_{12}}(r)}{E^{(0)}_{n'0} - E^{(0)}_{m2}} 
\frac{I_3 (n1 \rightarrow m2)}{I_3 (n1 \rightarrow n'0)} 
\langle {\bf S}_{12} \rangle_{1J1}\,.
\label{ST2}
\end{align}
Analogous results hold for the other potentials. The complete correction to the transition width coming from 
higher-order potentials is of the form
\bea
\bar{{\Gamma}}^{\textrm{h.o. potentials}}_{n^3P_J \rightarrow n^{\prime\,3}S_1 \,\gamma} = R^{S=1}_{nn'}(J) \,,
\\
\bar{{\Gamma}}^{\textrm{h.o. potentials}}_{n^1P_1 \rightarrow n^{\prime\,1}S_0 \,\gamma} = R^{S=0}_{nn'} \,,
\eea
where the general spin structure of $R^{S=1}_{nn'}(J)$ is 
$R^{S=1}_{nn'}(J) =  A_{nn'} + 2 B_{nn'} + C_{nn'}J(J+1) + D_{nn'}\langle {\bf S}_{12} \rangle_{1J1}$ 
($B_{nn'}$ is the correction coming from the spin-spin potential)
and the general spin structure of $R^{S=0}_{nn'}$ is 
$R^{S=0}_{nn'}=  A_{nn'} + 2C_{nn'}$. The specific values of the coefficients $A_{nn'}$, $B_{nn'}$, $C_{nn'}$ and $D_{nn'}$, 
which involve expressions similar to the ones displayed in Eqs. \eqref{ST1} and \eqref{ST2}, 
depend on the form of the potentials and will not be discussed further in this work.

\subsubsection{Corrections due to higher Fock states}
The quarkonium initial and final states may also get corrections from 
the coupling of the heavy quark-antiquark pair to other low-energy 
degrees of freedom. We call these corrections higher Fock state corrections. 

For strongly-coupled quarkonia, we argued that we can neglect couplings with 
other low-energy degrees of freedom, see Sec.~\ref{secpNRQCD}. In this case, we do not have 
new corrections coming from higher Fock states. 

\begin{figure}[t]
  \centering
  \epsfig{file=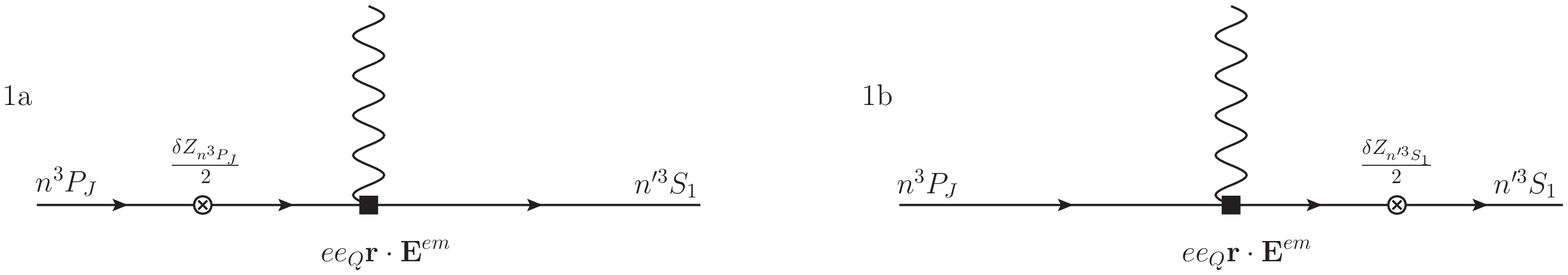,width=\linewidth,clip=}
  \epsfig{file=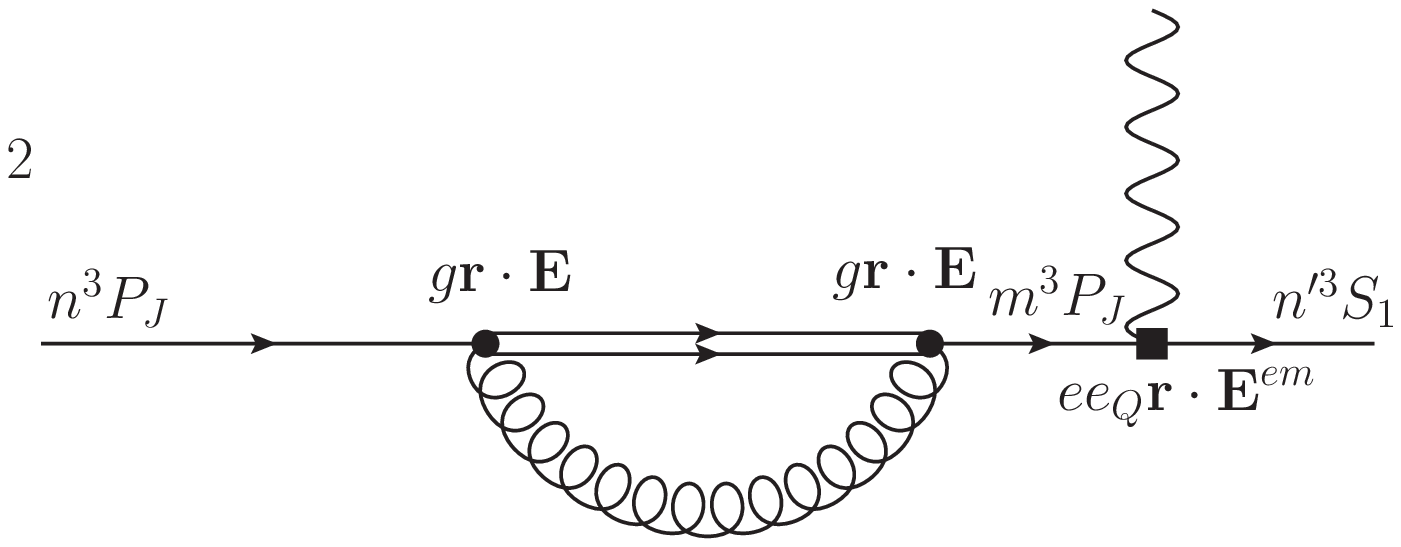,width=0.5\linewidth,clip=}
  \epsfig{file=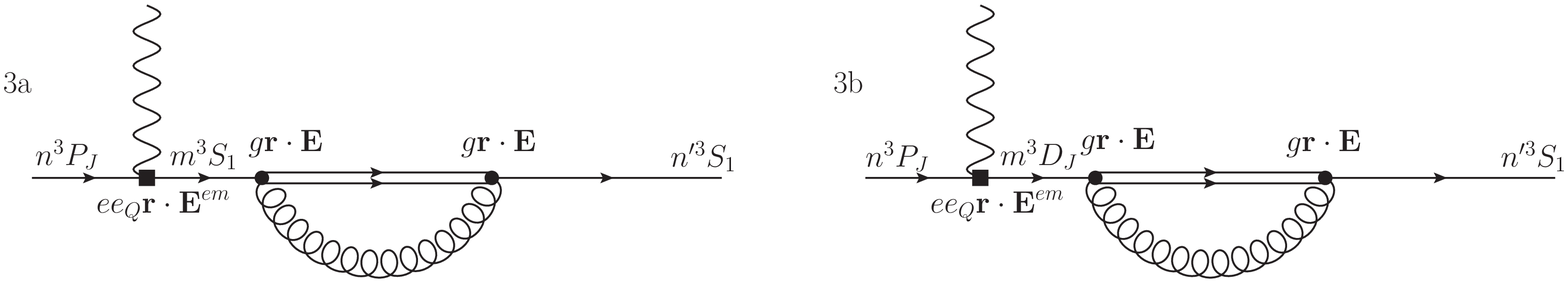,width=\linewidth,clip=}
  \epsfig{file=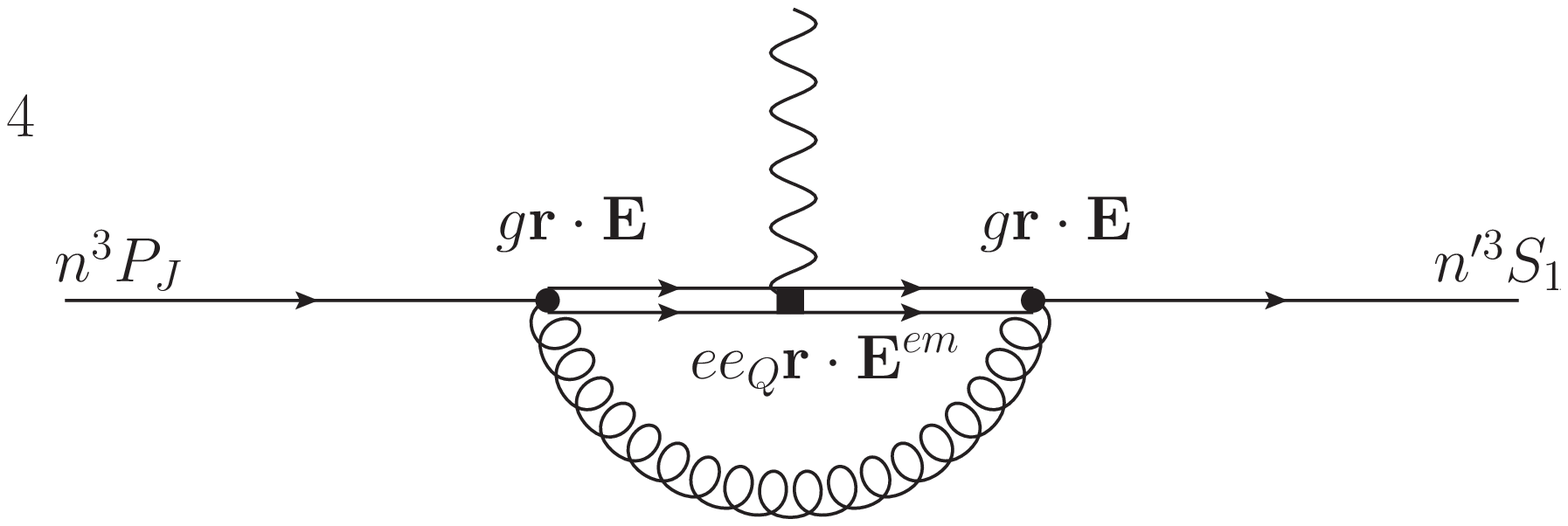,width=0.5 \linewidth,clip=}
  \caption{Color-octet contributions to the E1 transition $n^3P_J \rightarrow n^{\prime\,3}S_1 \,\gamma$ for 
  weakly-coupled states. The double line stands for an intermediate octet state.}
  \label{color-octet}
\end{figure}

For weakly-coupled quarkonia, we have to account for the coupling 
with low-energy gluons, see Eq. \eqref{Loctet}. A higher Fock state contributing 
at order $v^2$ is made of a gluon and a heavy quark-antiquark pair in a 
color-octet configuration. Moreover, photons may couple
to a quark-antiquark octet state through  \eqref{E1octet}. 
Again we consider the radiative transition $n^3P_J \rightarrow n^{\prime\,3}S_1 \,\gamma$, whose 
relevant diagrams at relative order $v^2$ are shown in Fig.~\ref{color-octet}.

The first two diagrams correspond to the normalization of the initial and final states, 
$(\delta Z _{n^3 P_J} + \delta Z _{n^{\prime\,3} S_1})/2$. 
They can be calculated from the self energy of the state, first derived in \cite{hep-ph/9907240}, 
which is given for a generic quarkonium $H$ by\footnote{
With respect to \cite{hep-ph/9907240}, a factor $1/N_c$ has been reabsorbed into the 
normalization of the vacuum state.}
\bea
\delta E _{H(\lambda)} &=& 
- \frac{i}{6} \int_{0}^{\infty} \!\!dt \;
\langle \textrm{vac} \lvert g E^{a\,i}({\bf R},t) \phi (t,0)^{\textrm{adj}}_{ab} g E^{b\,i}({\bf R},0) \lvert \textrm{vac} \rangle 
\nonumber\\
&&
\hspace{4cm}
\times
{}^{(0)}\langle H({\bf 0},\lambda) \lvert r^j e^{-i(h_O ^{(0)}-E_H ^{(0)}) t} r^j \lvert H({\bf 0},\lambda) \rangle^{(0)} 
\,,
\eea
where the Wilson line in the adjoint representation is 
\begin{equation}
\phi (t,0)^{\textrm{adj}}_{ab} = \left( e^{\displaystyle -ig \int_0^t dt' A_0 ({\bf R},t') ^{\textrm{adj}}} \right)_{ab} \, ,  
\end{equation}
and $h_O ^{(0)} \equiv -\bfnabla^2_r/m + V_O^{(0)}$.
Deriving the self energy with respect to the energy provides the state normalization
\bea
\delta Z _{H(\lambda)} =  \frac{\partial \delta E _{H(\lambda)}}{\partial E ^{(0)}_{H}} 
&=& 
\frac{1}{6} \int_{0}^{\infty} \!\!dt\,t \;
\langle \textrm{vac} \lvert g E^{a\,i}({\bf R},t) \phi (t,0)^{\textrm{adj}}_{ab} g E^{b\,i}({\bf R},0) \lvert \textrm{vac} \rangle 
\nonumber\\
&&
\hspace{2cm}
\times
{}^{(0)}\langle H({\bf 0},\lambda) \lvert r^j e^{-i(h_O ^{(0)}-E_H ^{(0)}) t} r^j \lvert H({\bf 0},\lambda) \rangle^{(0)} 
\,.
\eea
Thus, the contribution from the state normalizations to the amplitude reads
\bea
\mathcal{A}^{\textrm{Fig.~\ref{color-octet}},\;1}_{n^3P_J \rightarrow n^{\prime\,3}S_1 \,\gamma} &=& 
\frac{\mathcal{A}^{(0)}_{n^3 P_J \rightarrow n^{\prime\,3} S_1 \,\gamma}}{12}  
\int_{0}^{\infty} dt \, t \; 
\langle \textrm{vac} \lvert g E^{a\,i}({\bf R},t) \phi (t,0)^{\textrm{adj}}_{ab} g E^{b\,i}({\bf R},0) \lvert \textrm{vac} \rangle 
\nonumber\\
&& \hspace{-2.4cm}
\times \left[ {}^{(0)}\langle n^3 P_J \lvert r^j e^{-i(h_O ^{(0)}-E_{n1} ^{(0)}) t} r^j \lvert n^3 P_J \rangle^{(0)} 
+ {}^{(0)}\langle n^{\prime\,3} S_1 \lvert r^j e^{-i(h_O ^{(0)}-E_{n'0} ^{(0)}) t} r^j \lvert n^{\prime\,3} S_1 \rangle^{(0)} 
\right].
\label{81}
\eea
Diagram 2, the initial state correction, yields
\bea
\mathcal{A}^{\textrm{Fig.~\ref{color-octet}},\;2}_{n^3P_J \rightarrow n^{\prime\,3}S_1 \,\gamma}  &=& 
-  \frac{i}{6} \sum _{m \neq n} \frac{\mathcal{A}^{(0)}_{m^3 P_J \rightarrow n^{\prime\,3} S_1 \,\gamma}}{E_{n1} ^{(0)} - E _{m1}^{(0)}} 
\int_0^{\infty} dt \, 
\langle \textrm{vac} \lvert g E^{a\,i}({\bf R},t) \phi (t,0)^{\textrm{adj}}_{ab} g E^{b\,i}({\bf R},0) \lvert \textrm{vac} \rangle 
\nonumber\\
&& \hspace{-1cm}
\times
{}^{(0)}\langle m^3 P_J \lvert r^j e^{-i(h_O ^{(0)}-E_{n1} ^{(0)}) t} r^j \lvert n^3 P_J \rangle^{(0)} \,,
\eea
while diagrams 3a and 3b, the final state corrections, yield
\bea
\mathcal{A}^{\textrm{Fig.~\ref{color-octet}},\;3\textrm{a}}_{n^3P_J \rightarrow n^{\prime\,3}S_1 \,\gamma} &=& 
- \frac{i}{6} \sum _{m \neq n'} \frac{\mathcal{A}^{(0)}_{n^3 P_J \rightarrow m^3 S_1 \,\gamma}}{E_{n'0} ^{(0)} - E _{m0}^{(0)}} 
\int_0^{\infty} dt \, 
\langle \textrm{vac} \lvert g E^{a\,i}({\bf R},t) \phi (t,0)^{\textrm{adj}}_{ab} g E^{b\,i}({\bf R},0) \lvert \textrm{vac} \rangle 
\nonumber\\
&& \hspace{-1cm}
\times
{}^{(0)}\langle  n^{\prime\,3}S_1 \lvert r^j e^{-i(h_O ^{(0)}-E_{n'0} ^{(0)}) t} r^j \lvert   m^3S_1 \rangle^{(0)} 
\,,  
\\
\mathcal{A}^{\textrm{Fig.~\ref{color-octet}},\;3\textrm{b}}_{n^3P_J \rightarrow n^{\prime\,3}S_1 \,\gamma} &=& 0 \, .
\eea
$\mathcal{A}^{\textrm{Fig.~\ref{color-octet}},\;3\textrm{b}}$ vanishes, since 
$r^j \, \textrm{e}^{-i(h_O ^{(0)}-E_{n'0} ^{(0)})t} \, r^j$ is a scalar that cannot
change the angular momentum. The contribution from the last diagram,
where the photon is coupled to an intermediate octet state, reads
\bea
 \mathcal{A}^{\textrm{Fig.~\ref{color-octet}},\;4}_{n^3P_J \rightarrow n^{\prime\,3}S_1 \,\gamma} &=& 
- \frac{i}{6} \int_0^{\infty} dt \int_0^{t} dt' \, 
\langle \textrm{vac} \lvert g E^{a\,i}({\bf R},t) \phi (t,0)^{\textrm{adj}}_{ab} g E^{b\,i}({\bf R},0) \lvert \textrm{vac} \rangle 
\nonumber\\
&& \hspace{-1cm}
\times
{}^{(0)}\langle n^{\prime\,3} S_1 \lvert r^{j} e^{-i(h_O ^{(0)}-E_{n'0} ^{(0)}) (t-t')} 
{e e_Q k {\boldsymbol \epsilon}^\ast(\sigma) \cdot \bf r}  
e^{-i(h_O ^{(0)}-E_{n1} ^{(0)}) t'} r^{j} \lvert n^3 P_J \rangle^{(0)}.
\label{84}
\eea
According to the power counting of Sec.~\ref{secpNRQCD}, all amplitudes 
\eqref{81}-\eqref{84} contribute to relative order 
$1/(mv)^2 \times \lQ^2 \sim  v^2$ or $1/(mv)^2 \times \lQ^3 \times 1/(mv^2) \sim  v^2$, 
where the factor $1/(mv)^2$ comes from the two $r^j$ and we have assumed 
the chromoelectric correlator to scale like $\lQ^4\sim (mv^2)^4$.

It is noteworthy that, in contrast to M1 transitions \cite{hep-ph/0512369}, 
weak-coupling non-perturbative contributions do not cancel for E1 transitions. 
This is due to the fact that the leading electric dipole operator does not commute with the kinetic energy.
Hence, even in the weak coupling, E1 transitions are affected at order 
$v^2$ by non-perturbative contributions. These are encoded in the chromoelectric correlator 
$\langle \textrm{vac} \lvert g E^{a\,i}({\bf R},t) \phi (t,0)^{\textrm{adj}}_{ab} g E^{b\,i}({\bf R},0) \lvert \textrm{vac} \rangle$.
For $E \gg \lQ$, the correlator reduces to the gluon condensate, which factorizes;
however, this scale hierarchy is unlikely to be of relevance for E1 transitions.

\section{Results and comparison with the literature}
\label{secRes}
Now we can summarize our results and compare them with \cite{PRINT-84-0308}.  
The results are obtained by summing up all corrections calculated in
the previous section, which include contributions coming from E1 operators of relative order $v^2$, and initial 
and final state corrections.
The complete decay rates $n^3P_J \rightarrow n^{\prime\,3}S_1 \gamma$ and $n^1P_J \rightarrow n^{\prime\,1}S_0 \gamma$,
this last one obtained by leaving out spin-dependent contributions from the former expression, 
read to order $k_{\gamma}^3/m^2$ 
\bea
\Gamma_{n^3P_J \rightarrow n^{\prime\,3}S_1 \,\gamma} &=& 
\Gamma^{(0)}_{nn'} \left[ 1 + R_{nn'}^{S=1}(J) - \frac{k_{\gamma}}{6m} 
- \frac{k_{\gamma}^2}{60} \frac{I_5 (n1 \rightarrow n'0)}{I_3(n1 \rightarrow n'0)} 
\right.
\nonumber\\
&&
\hspace{-2.5cm}
\left.
+ \left(\frac{J(J+1)}{2} -2 \right) \left(-(1 +  \kappa_Q^{\em})\frac{k_{\gamma}}{2m}
+ \frac{1}{m^2}(1+2\kappa_Q^{\em})\frac{I_2'(n1 \rightarrow n'0)+ 2 I_1(n1 \rightarrow n'0)}{I_3(n1 \rightarrow n'0)}\right) \right] , 
\nonumber\\
\label{res1}\\
\Gamma_{n^1P_1 \rightarrow n^{\prime\,1}S_0 \,\gamma} &=& 
\Gamma^{(0)}_{nn'} \left[ 1  + R_{nn'}^{S=0} - \frac{k_{\gamma}}{6m} 
- \frac{k_{\gamma}^2}{60} \frac {I_5(n1 \rightarrow n'0)}{I_3(n1 \rightarrow n'0)} \right]  .
\label{res2}
\eea
$R_{nn'}^{S=1}(J)$ and $R_{nn'}^{S=0}$ are the initial and final state corrections; 
they include corrections coming from higher-order potentials, see Sec.~\ref{sechopot},  
and, in the case of weakly-coupled quarkonia, the color-octet contributions 
computed in Eqs. \eqref{81}-\eqref{84}.
We have kept terms proportional to the anomalous magnetic moment, 
$\kappa_Q^{\em}$, for the sake of the following comparison, 
though these terms are suppressed by $\als(m)$ and go beyond our accuracy.
Analogously, the expressions for the decay rates $n^3S_1 \rightarrow n^{\prime\,3}P_J \gamma$ and $n^1S_0 \rightarrow n^{\prime\,1}P_1 \gamma$ 
read
\bea
\Gamma_{n^3S_1 \rightarrow n^{\prime\,3}P_J \,\gamma} &=& 
\frac{2J+1}{3}\Gamma^{(0)}_{nn'} \left[ 1 + R_{nn'}^{S=1}(J) + \frac{k_{\gamma}}{6m} 
- \frac{k_{\gamma}^2}{60} \frac {I_5(n'1 \rightarrow n0)}{I_3(n'1 \rightarrow n0)} 
\right.
\nonumber\\
&&
\hspace{-2.5cm}
\left.
+ \left(\frac{J(J+1)}{2} -2 \right) \left((1 + \kappa_Q^\em) \frac{k_{\gamma}}{2m} 
+ \frac{1}{m^2}(1+2\kappa_Q^\em)\frac{I_2 '(n'1 \rightarrow n0)+ 2 I_1(n'1 \rightarrow n0) } {I_3(n'1 \rightarrow n0)}\right) \right] , 
\nonumber\\
\label{res3}\\
\Gamma_{n^1S_0 \rightarrow n^{\prime\,1}P_1 \,\gamma} &=& 
3 \Gamma^{(0)}_{nn'} \left[ 1 + R_{nn'}^{S=0} + \frac{k_{\gamma}}{6m} 
- \frac{k_{\gamma}^2}{60} \frac {I_5(n'1 \rightarrow n0)}{I_3(n'1 \rightarrow n0)} \right].
\label{res4}
\eea

The decay rates derived in \cite{PRINT-84-0308} read
\bea
\Gamma ^{\textrm{\cite{PRINT-84-0308}}}_{n^3P_J \rightarrow n^{\prime\,3}S_1 \,\gamma} &=& 
\Gamma^{(0)}_{nn'} \left[ 1 +  R_{nn'}^{S=1}(J) - \frac{k_{\gamma}^2}{10} 
\frac {I_5(n1 \rightarrow n'0)}{I_3(n1 \rightarrow n'0)} 
+ \frac{k_{\gamma}}{2m}\frac {I'_4(n1 \rightarrow n'0)}{I_3(n1 \rightarrow n'0)} 
+ \frac{k_{\gamma}}{m} 
\right.
\nonumber\\
&&
\hspace{2cm}
\left.
+ \kappa_Q^\em\frac{k_{\gamma}}{2m} \left(\frac{J(J+1)}{2} -2 \right) \right] , 
\label{res1G}\\
\Gamma ^{\textrm{\cite{PRINT-84-0308}}}_{n^3S_1 \rightarrow n^{\prime\,3}P_J \,\gamma} &=& 
\frac{2J+1}{3}\Gamma^{(0)}_{nn'} \left[ 1 +  R_{nn'}^{S=1}(J) - \frac{k_{\gamma}^2}{10} 
\frac {I_5(n'1 \rightarrow n0)}{I_3(n'1 \rightarrow n0)} 
- \frac{k_{\gamma}}{2m}\frac {I'_4(n'1 \rightarrow n0)}{I_3(n'1 \rightarrow n0)} 
- \frac{k_{\gamma}}{m} 
\right.
\nonumber\\
&&
\hspace{2cm}
\left.
- \kappa_Q^\em\frac{k_{\gamma}}{2m} \left(\frac{J(J+1)}{2} -2 \right) \right] .
\label{res3G}
\eea
The expressions appear different from the ones derived in this work because 
the basis of operators used in  \cite{PRINT-84-0308} is different from the one 
of pNRQCD. The two basis are however related by a field redefinition so that at the end the 
two results are equivalent. To see this explicitly at the level of the transition widths, 
consider the radial Schr\"odinger equation
\begin{equation}
E_{nL}R_{nL}= \left[- \frac{1}{m}\left(\frac{\partial ^2}{\partial r ^2} +\frac{2}{r} \frac{\partial}{\partial r} \right) 
+ \frac{L(L+1)}{mr^2} + V_S^{(0)}(r)\right] R_{nL} \,.
\end{equation}
From there, it follows (up to corrections of relative order $v^2$)
\begin{eqnarray}
k_{\gamma} I_5(n1 \rightarrow n'0) &=& \frac{1}{m}\left[6 I_4'(n1 \rightarrow n'0) + 14 I_3(n1 \rightarrow n'0)\right], 
\label{id1}\\
k_{\gamma} I_3(n1 \rightarrow n'0)  &=&  \frac{1}{m}\left[2 I_2'(n1 \rightarrow n'0) + 4 I_1(n1 \rightarrow n'0)\right],
\label{id2}\\
- k_{\gamma} I_5(n'1 \rightarrow n0) &=& \frac{1}{m}\left[6 I_4'(n'1 \rightarrow n0) + 14 I_3(n'1 \rightarrow n0)\right], 
\label{id3}\\
- k_{\gamma} I_3(n'1 \rightarrow n0) &=&  \frac{1}{m}\left[2 I_2'(n'1 \rightarrow n0) + 4 I_1(n'1 \rightarrow n0)\right]\,,
\label{id4}
\end{eqnarray}
where $k_{\gamma} = E^{(0)}_{n1}-E^{(0)}_{n'0} + {\cal O}(v^4)$ in the first two lines  
and $k_{\gamma} = E^{(0)}_{n0}-E^{(0)}_{n'1}+ {\cal O}(v^4)$ in the third and fourth line.  
Using the identities \eqref{id1}-\eqref{id4}, Eq. \eqref{res1} 
can be cast in the form of Eq. \eqref{res1G} and Eq. \eqref{res3} in the form of Eq. \eqref{res3G}.
Finally, a third way of presenting the same results, but in a more compact form, is 
\bea
\Gamma _{n^3P_J \rightarrow n^{\prime\,3}S_1 \,\gamma} &=& 
\Gamma^{(0)}_{nn'} \left[ 1 +  R_{nn'}^{S=1}(J) 
- \frac{k^2_\gamma}{60} \frac {I_5(n1 \rightarrow n'0)}{I_3(n1 \rightarrow n'0)} - \frac{k_{\gamma}}{6m} 
\right.
\nonumber\\
&&
\hspace{6cm}
\left.
+  \kappa_Q^\em \frac{k_{\gamma}}{2m} \left(\frac{J(J+1)}{2} -2 \right) \right] , 
\label{res1compact} 
\\
\Gamma _{n^3S_1 \rightarrow n^{\prime\,3}P_J \,\gamma} &=& 
\frac{2J+1}{3}\Gamma^{(0)}_{nn'} \left[ 1 +  R_{nn'}^{S=1}(J) 
- \frac{k^2_\gamma}{60} \frac {I_5(n'1 \rightarrow n0)}{I_3(n'1 \rightarrow n0)} + \frac{k_{\gamma}}{6m} 
\right.
\nonumber\\
&&
\hspace{6cm}
\left.
-  \kappa_Q^\em\frac{k_{\gamma}}{2m} \left(\frac{J(J+1)}{2} -2 \right) \right] . 
\label{res3compact}
\eea

In summary, the decay width $\Gamma_{n^3P_J \rightarrow n^{\prime\,3}S_1 \,\gamma}$ may be written up to order 
$k_{\gamma}^3/m^2$ in the equivalent ways \eqref{res1}, \eqref{res1G} or \eqref{res1compact},
the decay width $\Gamma_{n^3S_1 \rightarrow n^{\prime\,3}P_J \,\gamma}$ in the equivalent ways 
\eqref{res3}, \eqref{res3G} or \eqref{res3compact}, the decay width 
$\Gamma_{n^1P_1 \rightarrow n^{\prime\,1}S_0 \,\gamma}$ as in Eq. \eqref{res2} and 
the decay width  $\Gamma_{n^1S_0 \rightarrow n^{\prime\,1}P_1 \,\gamma}$ as in Eq. \eqref{res4}.
The obtained expressions are valid both for weakly-coupled and strongly-coupled quarkonia, 
the difference between the two cases being in the wave functions. 
Initial- and final-state wave functions affect crucially electric dipole transitions. 
The leading-order width, $\Gamma^{(0)}_{nn'}$, depends on the wave functions, and 
at higher orders the integrals $I_N$ and the initial- and final-state 
corrections $R_{nn'}^{S=1}(J)$ and $R_{nn'}^{S=0}$ depend on the wave functions.

In the case of weakly-coupled quarkonia, the wave functions are Coulombic, which implies 
that $\Gamma^{(0)}_{nn'}$, $I_N$ and the initial- and final-state corrections due to higher-order 
potentials may be calculated in perturbation theory. The relevant potentials are those 
listed in Eqs. \eqref{m2pot}-\eqref{pertpot}. Weakly-coupled initial and final states 
get also corrections due to color-octet quark antiquark states coupled to low-energy gluons, 
which are parametrically of the same order as the other corrections.  
Color-octet corrections are given by Eqs. \eqref{81}-\eqref{84} and depend on the 
correlator of two chromoelectric fields, which is a non-perturbative quantity.
Therefore, at relative order $v^2$ even E1 transitions of weakly-coupled quarkonia are affected by non-perturbative 
corrections.

In the case of strongly-coupled quarkonia, the potentials and, hence, the wave functions are non-Coulombic.
The integrals $I_N$ and the wave-function corrections due to higher-order potentials, which are 
encoded in the coefficients $A_{nn'}$, $B_{nn'}$, $C_{nn'}$ and $D_{nn'}$ defined in Sec.~\ref{sechopot}, 
are non-perturbative parameters. These non-perturbative parameters may be either derived 
from the quarkonium potentials evaluated on the lattice or fitted to the data.
For strongly-coupled quarkonia, there are no relevant higher Fock state contributions to the initial and 
final states to be included.

\section{Conclusions and Outlook}
\label{secCon}
The paper completes the analysis of radiative transitions in an 
EFT framework initiated in \cite{hep-ph/0512369} with the study of M1 transitions.
The EFTs are NRQCD and pNRQCD.

The paper deals with E1 transitions, which are studied at relative order 
$v^2$, corresponding to order $k_{\gamma}^3/m^2$ in the transition width.
All the relevant operators of pNRQCD are listed in Eq. \eqref{LagE1}.
The matching, performed in Sec.~\ref{secMatch}, shows that, if charm-loop effects are 
neglected, these operators do not get corrections from the momentum-transfer scale 
and keep the value inherited from NRQCD to all orders in perturbation theory 
and non-perturbatively. Charm-loop effects may be treated perturbatively and affect 
the matching beyond our accuracy. This non-obvious outcome may be considered the main 
result of the paper.

As a consequence of the exact matching, we could provide the transition widths 
$\Gamma_{n^3P_J \rightarrow n^{\prime\,3}S_1 \,\gamma}$,  $\Gamma_{n^3S_1 \rightarrow n^{\prime\,3}P_J \,\gamma}$, 
$\Gamma_{n^1P_1 \rightarrow n^{\prime\,1}S_0 \,\gamma}$ and $\Gamma_{n^1S_0 \rightarrow n^{\prime\,1}P_1 \,\gamma}$ 
up to order $k_{\gamma}^3/m^2$ for both weakly- and strongly-coupled quarkonia.
Weakly-coupled quarkonia are those bound by a Coulombic potential, possible states 
being the quarkonium ground states; strongly-coupled quarkonia are those bound by a 
non-perturbative potential, which eventually becomes confining in the long range. 
Strongly-coupled quarkonia are likely all states above the ground state.
The transition widths have the same expressions for weakly- and strongly-coupled quarkonia, 
the only difference lying in the wave functions and ultimately in the potentials. 
Weakly-coupled states also get corrections from intermediate quark-antiquark color octet states.
The final expressions of the transition widths are listed in Sec.~\ref{secRes}.
Many alternative expressions for the widths are possible, all of them 
equivalent at order $k_{\gamma}^3/m^2$. We have listed some of them 
in the case of $n^3P_J \rightarrow n^{\prime\,3}S_1 \,\gamma$ and $n^3S_1 \rightarrow n^{\prime\,3}P_J \,\gamma$ 
transitions, equations \eqref{res1compact} and \eqref{res3compact} providing the most compact expressions.

The expressions for the widths agree, with some specifications, with 
the expressions obtained in \cite{PRINT-84-0308} by reducing some covariant two-particle 
bound state equation. The specifications are the followings. The expressions for the transition 
widths are valid up to relative order $v^2$. At this order the anomalous 
magnetic moment of the quark, $\kappa_Q^{\em}$, does not need to be included. 
If the anomalous magnetic moment is included, its expression is \eqref{kQ}. 
This amounts to a small positive quantity of order $\als(m)$. 
No large non-perturbative correction affects $\kappa_Q^{\em}$, 
as sometimes required in phenomenological treatments. 
According to a commonly used power counting, wave-function corrections of relative order $v^2$ are 
induced by the potentials listed in Eqs. \eqref{m2pot}-\eqref{SD_potentials}.
Typically, corrections induced by the $1/m$ potential, $V^{(1)}$, have been neglected in 
the past, for the potential does not show up at tree level. 
For weakly-coupled quarkonia, non-perturbative corrections induced 
by low-energy gluons coupled to color octet quark-antiquark states 
have to be included as well. These corrections have been computed here for the first 
time and have not been included in any earlier treatment although they contribute 
at relative order $v^2$.

Relativistic corrections to E1 transitions have, in some respects,  
opposite characteristics to the ones to M1 transitions. 
Allowed M1 transitions between quarkonium ground states 
can be described at relative order $v^2$ entirely in perturbation 
theory. In contrast, E1 transitions, even between weakly-coupled quarkonia, 
require at relative order $v^2$ a non-perturbative input. The reason is that, 
while the magnetic dipole operator commutes with the kinetic energy, leading 
eventually to the cancellation of the octet contributions, the electric 
dipole operator does not. M1 transitions between strongly-coupled 
quarkonia require at relative order $v^2$ the non-perturbative matching 
of a yet unknown $1/m^3$ operator. In contrast, E1 transitions between strongly-coupled quarkonia involve 
at most $1/m^2$ operators, which are exactly known. Hence, a first principle calculation 
of E1 transitions at relative order $v^2$ is at present possible for all quarkonium states.
Clearly, in the case of strongly coupled quarkonia, this requires parameterizing 
the lattice quarkonium potentials and solving the corresponding Schr\"odinger 
equation.

Future applications of the present work include the numerical determination 
of the E1 transition widths between all $S$- and $P$-wave quarkonium 
states from the expressions given in  Sec.~\ref{secRes}.
A consistent determination would require parameterizing the long distance 
behaviour of the quarkonium potentials as evaluated on the lattice 
(for recent lattice results see, for instance, \cite{Koma:2006si,Koma:2006fw,Koma:2007jq}), 
and matching it with the known short distance behaviour \cite{Piotr,Laschka:2011zr}, 
solving the corresponding Schr\"odinger equation and finally evaluating the integrals 
$I_N$ and the wave-functions corrections due to higher-order potentials. 
For weakly-coupled quarkonia, a parameterization of the chromoelectric field 
correlator would also be necessary.

Finally, we mention that the EFT approach for quarkonium radiative transitions 
discussed here can be translated to other systems beyond QCD. 
For instance, one could study atomic dipole transitions or 
dipole transitions in quirkonium, which is a candidate for dark matter \cite{arXiv:0909.2034}. 
The coupling constants of these systems are small, making them suited for a 
perturbative treatment.

\section*{Acknowledgments}
We thank Hector Martinez for collaboration at an early stage of this work.
N.B and A.V. acknowledge financial support from the DFG cluster of excellence ``Origin and structure of the universe''
(\href{http://www.universe-cluster.de}{www.universe-cluster.de}).

\newpage

\appendix

\section{Matching of M1 operators at $\mathcal{O}(1/m^3)$} 
Three  $1/m^3$  operators contribute to M1 transitions at relative order  $v^2$. 
One is the operator of Eq. \eqref{m3operator}, the other two are 
\begin{eqnarray}
&& \frac{1}{4 m^3} 
\; V^{\frac{\nabla_r^2 \, \sigma\cdot B}{m^3}}
\; \left\{ {\rm S}^\dagger , \bfsigma \cdot e e_Q {\bf B}^{\em} \right\} \bfnabla_r^2 {\rm S}\,, 
\label{app1}
\end{eqnarray}
and 
\begin{eqnarray}
&& \frac{1}{4 m^3} 
\; V^{\frac{(\nabla_r\cdot\sigma)\, (\nabla_r\cdot B)}{m^3}}
\; \left\{ {\rm S}^\dagger , \bfsigma^i \, e e_Q {\bf B}^{\em\,j} \right\} \bfnabla_r^i\bfnabla_r^j {\rm S}\,.
\label{app2}
\end{eqnarray}
An even number of momenta, $-i\bfnabla_r$, is required by time-reversal symmetry;
our analysis shows that other possible operators with two derivatives 
do not get contributions from the matching.

\begin{figure}[ht]
  \centering
  \epsfig{file=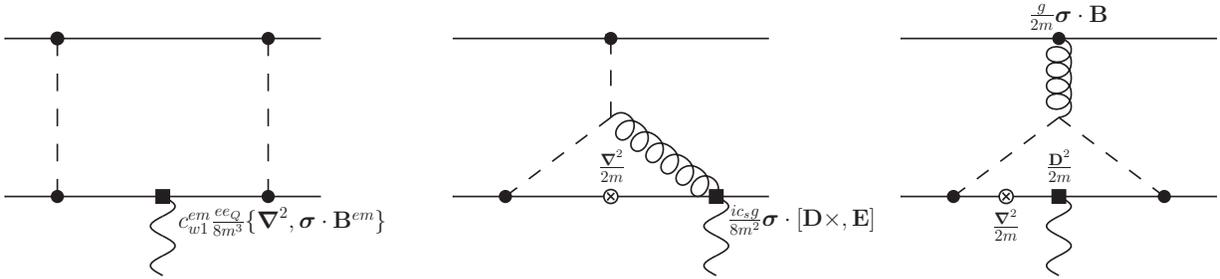,width=\linewidth,clip=}
  \caption{Some diagrams contributing to the matching at order $1/m^3$.}
  \label{m3diagrams}
\end{figure}

We focus on the operators \eqref{app1} and \eqref{app2}.
They come from matching spin-dependent NRQCD amplitudes that contain operators 
with at least two spatial derivatives (see  Fig.~\ref{m3diagrams}).
We have the following cases. 
 
1. Amplitudes that are made of a $\mathcal{O}(1/m^3)$ operator of 
NRQCD may couple to longitudinal photons, to photons embedded in 
covariant derivatives or to the magnetic fields of the operators.
The first case is excluded, for only transverse photons 
contribute to the operators \eqref{app1} and \eqref{app2}.
The second case does not contribute to the operators \eqref{app1} and \eqref{app2},  
for spin-dependent $\mathcal{O}(1/m^3)$ operators of NRQCD contain at most 
two covariant derivatives. The third case involves the NRQCD operators
\begin{eqnarray}
&& \frac{c^\em_{W1}}{8 m^3} \psi^\dagger \{ {\bf D}^2, \bfsigma \cdot e e_Q {\bf B}^{\em} \}\psi
- \frac{c^\em_{W2}}{4 m^3} \psi^\dagger {\bf D}^i \bfsigma \cdot e e_Q {\bf B}^{\em} {\bf D}^i \psi\,,
\\
&& \frac{c^\em_{p^\prime p}}{ 8 m^3} 
\psi^\dagger \left( \bfsigma\cdot{\bf D} \; e e_Q {\bf B}^\em \cdot {\bf D}
+ {\bf D} \cdot e e_Q {\bf B}^\em \;  \bfsigma\cdot{\bf D} \right) \psi\,,
\end{eqnarray} 
and the corresponding c.c. ones, 
with $c^\em_{W1}-c^\em_{W2} = 1$ and $c_{p^\prime p}^{\em} = {\cal O}(\als)$ \cite{hep-ph/0512369,hep-ph/9701294}. 
For the purpose of matching the operators \eqref{app1} and \eqref{app2}, the covariant derivatives 
may be replaced by simple derivatives that do not act on the gluon fields.  
We are, therefore, in the situation discussed in Sec.~\ref{secreducible}, paragraph {\it{(iv)}}, and 
Eq. \eqref{coem} leads to 
\begin{equation}
V^{\frac{\nabla_r^2 \, \sigma\cdot B}{m^3}}=1, \qquad 
V^{\frac{(\nabla_r\cdot\sigma)\, (\nabla_r\cdot B)}{m^3}} = c_{p^\prime p}^{\em}\,.
\label{app3}
\end{equation}
These relations are exact because, as we are going to detail in the 
following, NRQCD amplitudes with insertions of $1/m$ and $1/m^2$ operators
factorize at relative order $v^2$ into contributions to $1/m$ and
$1/m^2$ pNRQCD operators. 

2. In diagrams made of a kinetic energy and a chromoelectric
spin-orbit operator, the photon is coupled to one of the covariant
derivatives. If it is coupled to the covariant derivative of the
chromoelectric spin-orbit operator, then the two derivatives come from
the kinetic energy operator, which factorizes. If it is coupled to the
kinetic energy operator, then the photon field multiplies a
derivative, the operator commutes with the gluons and factorizes.

3. The situation is similar for diagrams with a kinetic energy and an 
electric spin-orbit operator insertion.

4. In diagrams made of two kinetic energy and a chromomagnetic dipole
operator, the photon is coupled to one of the kinetic energy
operators. If the photon field multiplies a derivative, then either
the operator commutes with the gluons and factorizes or it does not
and then the two derivatives come from the other kinetic energy
operator, which factorizes. If the photon field multiplies a gluon
field then the two derivatives come from the other kinetic energy
operator, which factorizes.

5. The kinetic energy operator also factorizes in diagrams with 
a magnetic and a chromomagnetic dipole operator.

6. Finally, in diagrams made of two kinetic energy and a magnetic dipole operator, 
the magnetic dipole operator commutes at relative order $v^2$ with the kinetic energy and factorizes.

We conclude therefore that \eqref{app3} holds to any order of the matching, while 
only the Wilson coefficient $V^{\frac{\sigma\cdot B}{m^3}}$ of (\ref{m3operator}), 
the only operator without derivatives, may possibly get corrections from the matching 
beyond tree level.


\begin{thebibliography}{99}
%\cite{Brambilla:2004wf}
\bibitem{Brambilla:2004wf}
  N.~Brambilla {\it et al.},
  CERN-2005-005, (CERN, Geneva, 2005)
  %``Heavy quarkonium physics,''
  [arXiv:hep-ph/0412158].
  %%CITATION = HEP-PH 0412158;%%

%\cite{arXiv:1010.5827}
\bibitem{arXiv:1010.5827} 
  N.~Brambilla, S.~Eidelman, B.~K.~Heltsley, R.~Vogt, G.~T.~Bodwin, E.~Eichten, A.~D.~Frawley and A.~B.~Meyer {\it et al.},
  %``Heavy quarkonium: progress, puzzles, and opportunities,''
  Eur.\ Phys.\ J.\ C\ {\bf 71}, 1534  (2011)
  [arXiv:1010.5827 [hep-ph]].
  %%CITATION = EPHJA,C71,1534;%%

%\cite{arXiv:1002.0501}
\bibitem{arXiv:1002.0501} 
  M.~Ablikim {\it et al.} [BESIII Collaboration],
  %``Measurements of h_c(^1P_1) in psi' Decays,''
  Phys.\ Rev.\ Lett.\ \ {\bf 104}, 132002  (2010)
  [arXiv:1002.0501 [hep-ex]].
  %%CITATION = PRLTA,104,132002;%%

%\cite{arXiv:0903.1124}
\bibitem{arXiv:0903.1124} 
  B.~Aubert {\it et al.} [BABAR Collaboration],
  %``Evidence for the eta(b)(1S) Meson in Radiative Upsilon(2S) Decay,''
  Phys.\ Rev.\ Lett.\ \ {\bf 103}, 161801  (2009)
  [arXiv:0903.1124 [hep-ex]].
  %%CITATION = PRLTA,103,161801;%%

%\cite{arXiv:0909.5474}
\bibitem{arXiv:0909.5474} 
  G.~Bonvicini {\it et al.} [CLEO Collaboration],
  %``Measurement of the eta(b)(1S) mass and the branching fraction for Upsilon(3S) ---> gamma eta(b)(1S),''
  Phys.\ Rev.\ D\ {\bf 81}, 031104  (2010)
  [arXiv:0909.5474 [hep-ex]].
  %%CITATION = PHRVA,D81,031104;%%

%\cite{arXiv:1102.4565}
\bibitem{arXiv:1102.4565} 
  J.~P.~Lees {\it et al.}  [The BABAR Collaboration],
  %``Evidence for the $h_b(1P)$ meson in the decay $\Upsilon(3S) \to \pi^0 h_b(1P)$,''
  Phys.\ Rev.\ D {\bf 84}, 091101 (2011)
  [arXiv:1102.4565 [hep-ex]];
  %%CITATION = ARXIV:1102.4565;%%
  R.~Mizuk  [BELLE Collaboration], talk at the QWG workshop, GSI, (2011).

%\cite{arXiv:1012.0589}
\bibitem{arXiv:1012.0589} 
  M.~Kornicer {\it et al.} [CLEO Collaboration],
  %``Measurements of branching fractions for electromagnetic transitions involving the $\chi_{bJ}(1P)$ states,''
  Phys.\ Rev.\ D\ {\bf 83}, 054003  (2011)
  [arXiv:1012.0589 [hep-ex]].
  %%CITATION = PHRVA,D83,054003;%%

%\cite{arXiv:1104.5254}
\bibitem{arXiv:1104.5254} 
  J.~P.~Lees {\it et al.} [BABAR Collaboration],
  %``A Study of radiative bottomonium transitions using converted photons,''
  Phys.\ Rev.\ D\ {\bf 84}, 072002  (2011)
  [Phys.\ Rev.\ D\ {\bf 84}, 099901  (2011)]
  [arXiv:1104.5254 [hep-ex]].
  %%CITATION = PHRVA,D84,072002;%%

%\cite{FERMILAB-PUB-10-665-PPD}
\bibitem{FERMILAB-PUB-10-665-PPD} 
  K.~Nakamura {\it et al.} [Particle Data Group Collaboration],
  %``Review of particle physics,''
  J.\ Phys.\ GG\ {\bf 37}, 075021  (2010).
  %%CITATION = JPHGB,G37,075021;%%

%\cite{PRINT-84-0308}
\bibitem{PRINT-84-0308} 
  H.~Grotch, D.~A.~Owen and K.~J.~Sebastian,
  %``Relativistic Corrections To Radiative Transitions And Spectra Of Quarkonia,''
  Phys.\ Rev.\ D\ {\bf 30}, 1924  (1984).
  %%CITATION = PHRVA,D30,1924;%%  

%\cite{hep-ph/0701208}
\bibitem{hep-ph/0701208} 
  E.~Eichten, S.~Godfrey, H.~Mahlke and J.~L.~Rosner,
  %``Quarkonia and their transitions,''
  Rev.\ Mod.\ Phys.\ \ {\bf 80}, 1161  (2008)
  [hep-ph/0701208].
  %%CITATION = RMPHA,80,1161;%%

%\cite{hep-ph/0410047}
\bibitem{hep-ph/0410047} 
  N.~Brambilla, A.~Pineda, J.~Soto and A.~Vairo,
  %``Effective field theories for heavy quarkonium,''
  Rev.\ Mod.\ Phys.\ \ {\bf 77}, 1423  (2005)
  [hep-ph/0410047].
  %%CITATION = RMPHA,77,1423;%%

%\cite{hep-ph/0512369}
\bibitem{hep-ph/0512369} 
  N.~Brambilla, Y.~Jia and A.~Vairo,
  %``Model-independent study of magnetic dipole transitions in quarkonium,''
  Phys.\ Rev.\ D\ {\bf 73}, 054005  (2006)
  [hep-ph/0512369].
  %%CITATION = PHRVA,D73,054005;%%

\bibitem{Piotr}  
P.~Pietrulewicz, 
{\it Effective field theory for electromagnetic transitions of heavy quarkonium},
Diploma Thesis (Munich, 2011).

%\cite{CLNS-85/641}
\bibitem{CLNS-85/641} 
  W.~E.~Caswell and G.~P.~Lepage,
  %``Effective Lagrangians for Bound State Problems in QED, QCD, and Other Field Theories,''
  Phys.\ Lett.\ B\ {\bf 167}, 437  (1986).
  %%CITATION = PHLTA,B167,437;%%

%\cite{hep-ph/9407339}
\bibitem{hep-ph/9407339} 
  G.~T.~Bodwin, E.~Braaten and G.~P.~Lepage,
  %``Rigorous QCD analysis of inclusive annihilation and production of heavy quarkonium,''
  Phys.\ Rev.\ D\ {\bf 51}, 1125  (1995)
  [Erratum-ibid.\ D\ {\bf 55}, 5853  (1997)]
  [hep-ph/9407339].
  %%CITATION = PHRVA,D51,1125;%%

%\cite{hep-ph/9707481}
\bibitem{hep-ph/9707481} 
  A.~Pineda and J.~Soto,
  %``Effective field theory for ultrasoft momenta in NRQCD and NRQED,''
  Nucl.\ Phys.\ Proc.\ Suppl.\ \ {\bf 64}, 428  (1998)
  [hep-ph/9707481].
  %%CITATION = NUPHZ,64,428;%%

%\cite{hep-ph/9907240}
\bibitem{hep-ph/9907240} 
  N.~Brambilla, A.~Pineda, J.~Soto and A.~Vairo,
  %``Potential NRQCD: An Effective theory for heavy quarkonium,''
  Nucl.\ Phys.\ B\ {\bf 566}, 275  (2000)
  [hep-ph/9907240].
  %%CITATION = NUPHA,B566,275;%%

%\cite{hep-ph/9205228}
\bibitem{hep-ph/9205228} 
  M.~E.~Luke and A.~V.~Manohar,
  %``Reparametrization invariance constraints on heavy particle effective field theories,''
  Phys.\ Lett.\ B\ {\bf 286}, 348  (1992)
  [hep-ph/9205228].
  %%CITATION = PHLTA,B286,348;%%

%\cite{hep-ph/9701294}
\bibitem{hep-ph/9701294} 
  A.~V.~Manohar,
  %``The HQET / NRQCD Lagrangian to order alpha / m-3,''
  Phys.\ Rev.\ D\ {\bf 56}, 230  (1997)
  [hep-ph/9701294].
  %%CITATION = PHRVA,D56,230;%%

%\cite{Brambilla:2009bi}
\bibitem{Brambilla:2009bi}
  N.~Brambilla, X.~Garcia i Tormo, J.~Soto and A.~Vairo,
  %``The QCD static energy at NNNLL,''
  Phys.\ Rev.\  D {\bf 80} (2009) 034016
  [arXiv:0906.1390 [hep-ph]].
  %%CITATION = PHRVA,D80,034016;%%

%\cite{hep-ph/0002250}
\bibitem{hep-ph/0002250} 
  N.~Brambilla, A.~Pineda, J.~Soto and A.~Vairo,
  %``The QCD potential at O(1/m),''
  Phys.\ Rev.\ D\ {\bf 63}, 014023  (2001)
  [hep-ph/0002250].
  %%CITATION = PHRVA,D63,014023;%%

%\cite{hep-ph/0009145}
\bibitem{hep-ph/0009145} 
  A.~Pineda and A.~Vairo,
  %``The QCD potential at O (1 / $m^{2)}$ : Complete spin dependent and spin independent result,''
  Phys.\ Rev.\ D\ {\bf 63}, 054007  (2001)
  [Erratum-ibid.\ D\ {\bf 64}, 039902  (2001)]
  [hep-ph/0009145].
  %%CITATION = PHRVA,D63,054007;%%

%\cite{hep-ph/0005066}
\bibitem{hep-ph/0005066} 
  D.~Eiras and J.~Soto,
  %``Light fermion finite mass effects in non-relativistic bound states,''
  Phys.\ Lett.\ B\ {\bf 491}, 101  (2000)
  [hep-ph/0005066].
  %%CITATION = PHLTA,B491,101;%%

%\cite{hep-ph/0108084}
\bibitem{hep-ph/0108084} 
  N.~Brambilla, Y.~Sumino and A.~Vairo,
  %``Quarkonium spectroscopy and perturbative QCD: Massive quark loop effects,''
  Phys.\ Rev.\ D\ {\bf 65}, 034001  (2002)
  [hep-ph/0108084].
  %%CITATION = PHRVA,D65,034001;%%

%\cite{hep-ph/0306107}
\bibitem{hep-ph/0306107} 
  N.~Brambilla, D.~Gromes and A.~Vairo,
  %``Poincare invariance constraints on NRQCD and potential NRQCD,''
  Phys.\ Lett.\ B\ {\bf 576}, 314  (2003)
  [hep-ph/0306107].
  %%CITATION = PHLTA,B576,314;%%

%\cite{arXiv:1007.4541}
\bibitem{arXiv:1007.4541} 
  Y.~Jia, W.~-L.~Sang and J.~Xu,
  %``Is the $J^P=2^-$ assignment for the $X(3872)$ compatible with the radiative transition data?,''
  arXiv:1007.4541 [hep-ph].
  %%CITATION = ARXIV:1007.4541;%%

%\cite{Eichten:1978tg}
\bibitem{Eichten:1978tg} 
  E.~Eichten, K.~Gottfried, T.~Kinoshita, K.~D.~Lane and T.~-M.~Yan,
  %``Charmonium: The Model,''
  Phys.\ Rev.\ D {\bf 17}, 3090 (1978)
  [Erratum-ibid.\ D {\bf 21}, 313 (1980)].
  %%CITATION = PHRVA,D17,3090;%%

%\cite{hep-ph/9910238}
\bibitem{hep-ph/9910238} 
  N.~Brambilla, A.~Pineda, J.~Soto and A.~Vairo,
  %``The Heavy quarkonium spectrum at order m alpha**5(s) l n alpha(s),''
  Phys.\ Lett.\ B\ {\bf 470}, 215  (1999)
  [hep-ph/9910238].
  %%CITATION = PHLTA,B470,215;%%

\bibitem{BS}
H.~A.~Bethe and E.~E.~Salpeter, {\it Quantum Mechanics of One and Two-Electron 
Atoms} (Plenum, New York, 1977).

%\cite{Koma:2006si}
\bibitem{Koma:2006si}
  Y.~Koma, M.~Koma and H.~Wittig,
  %``Nonperturbative determination of the QCD potential at $O(1/m)$,''
  Phys.\ Rev.\ Lett.\  {\bf 97}, 122003 (2006) 
  [arXiv:hep-lat/0607009].
  %%CITATION = PRLTA,97,122003;%%

%\cite{Koma:2006fw}
\bibitem{Koma:2006fw}
  Y.~Koma and M.~Koma,
  %``Spin-dependent potentials from lattice QCD,''
  Nucl.\ Phys.\  B {\bf 769}, 79 (2007) 
  [arXiv:hep-lat/0609078].
  %%CITATION = NUPHA,B769,79;%%

%\cite{Koma:2007jq}
\bibitem{Koma:2007jq}
  Y.~Koma, M.~Koma and H.~Wittig,
  %``Relativistic corrections to the static potential at $O(1/m)$ and $O(1/m^2)$,''
  PoS {\bf LAT2007}, 111 (2007) 
  [arXiv:0711.2322 [hep-lat]].
  %%CITATION = POSCI,LAT2007,111;%%

%\cite{Laschka:2011zr}
\bibitem{Laschka:2011zr} 
  A.~Laschka, N.~Kaiser and W.~Weise,
  %``Quark-antiquark potential to order 1/m and heavy quark masses,''
  Phys.\ Rev.\ D {\bf 83}, 094002 (2011)
  [arXiv:1102.0945 [hep-ph]].
  %%CITATION = ARXIV:1102.0945;%%

%\cite{arXiv:0909.2034}
\bibitem{arXiv:0909.2034} 
  G.~D.~Kribs, T.~S.~Roy, J.~Terning and K.~M.~Zurek,
  %``Quirky Composite Dark Matter,''
  Phys.\ Rev.\ D\ {\bf 81}, 095001  (2010)
  [arXiv:0909.2034 [hep-ph]].
  %%CITATION = PHRVA,D81,095001;%%
\end{thebibliography}
\end{document}